\documentclass[letterpaper,11pt]{article}
\pdfoutput=1
\usepackage{jheppub_mod}
\usepackage{graphicx, amsmath, amssymb, amstext,xcolor,txfonts}
\usepackage{relsize, float, multirow}
\usepackage{array, epstopdf, hyperref}
\usepackage{caption, subcaption, mathtools}
\usepackage{newtxmath,ulem}
\usepackage{multicol}
\definecolor{seagreen}{rgb}{0.18, 0.55, 0.34}

\title{ The role of the galaxy stellar mass function in determining the cosmological distribution of astrophysical transients with applications to fast radio bursts and merging binary black holes }

\author[a,b,1]{Sandeep Kumar Acharya\note{Corresponding author.}}
\affiliation[a]{Indian Institute of Astrophysics, Koramangala II Block, Bengaluru 560 034, India}
\affiliation[b]{Astrophysics Research Center of the Open University, The Open University of Israel, Ra'anana, Israel}
\emailAdd{sandeepkumaracharya92@gmail.com}
\date{\today}

\abstract{The cosmological distribution and formation rate of compact astrophysical objects such as fast radio bursts (FRBs) are typically assumed to be proportional to a linear combination of cosmological star formation rate and stellar mass. In the literature, a template for star formation rate, which is just a function of redshift, is typically used. In this work, we point out the importance of galaxy stellar mass function which  captures the host galaxy information of observed FRBs as well as the redshift evolution of galaxy stellar mass. 
Using this information \color{black} and taking the stellar mass distribution of a sample of localized FRBs at face value, \color{black} we find that FRB formation efficiency per stellar mass may have to be more efficient (by a factor of $\approx 3$) than previously calculated, in order to reproduce the observed volumetric rate of FRBs at $z=0$. 
We show that cosmological population studies of FRBs have to include host galaxy information along with its redshift evolution in order to obtain unbiased results. This consideration is also applicable to other transients, e.g. gamma-ray bursts and merging binary black hole events. We show that our approach may open up the possibility to \color{black} distinguish between different scenarios of merging binary black holes formation \color{black} with a detection of few thousand gravitational wave events.  }
\notoc

\notoc
\begin{document}
\maketitle
\newpage

%----------------------------------------------
\section{Introduction}
%----------------------------------------------
Fast radio bursts (FRBs) are radio transients, typically with milliseconds durations and originating from extragalactic distances. Till date thousands of distinct sources have been detected \citep{Chime_catalog,James2022,Sharma2024,Connor2024,Shannon2024}. So far, only about about a hundred of them have been localized to host galaxies. Localizing FRBs potentially allows us to study the environment which produces these sources. The aim is to find a correlation between the observed burst properties and the environment where these FRB sources reside. For example, there are various apparent differences between the properties of sources that are seen as repeaters and those that have thus far been detected only once, including spectro-temporal differences, polarization, distances and energetics (see \cite{BK2025B} for a discussion). While these properties might be due to selection effects, and the underlying FRB population appears to be consistent with a single underlying population of sources underlying repeaters and apparent non-repeaters \citep{BK2025,BK2025B} and no significant difference has been so far observed in their host galaxy properties \cite{Sharma2024,2025arXiv250215566L}, it is still important to critically examine the potential existence of FRB sub-populations by studying detailed host galaxy properties. In general, this idea has proven useful in transient studies. As an example, the environments of Type Ia supernovae are very distinct from core-collapse supernovae. Similarly, the environments of long and short GRBs are very different. In both cases, these differences helped to sub-divide the population into astrophysical distinct channels.

FRB formation is expected to be related to the star formation rate due to the various lines of evidence connecting them with magnetars \citep{PP2010,KLB2017,Wadiasingh2019,Cheng+20,BK2023,Totani2023}. This idea was validated with the association of a FRB-like burst with a known galactic magnetar \citep{CHIME2020,STARE2020}. Furthermore, several FRBs have been localized to star-forming galaxies \citep{Bhandari2022,Gordon2023_1}. 
However, there are several instances of FRBs localized to quiescent galaxies with significantly lower star formation activities \citep{Bannister2019,Ravi2019,Shah2025,Eftekhari2025} though they seem to constitute $\lesssim 10\%$ of whole population. Moreover, there can be significant star formation activity in a small, localized patch in a host galaxy with much less activity elsewhere which would make such a galaxy identified as a quiescent type. Indeed, the authors of \cite{Chen2025} reported higher star formation at the location of FRB20190520B than elsewhere inside its host galaxy. Instead, FRB20200120E has been localized to an old environment - a globular cluster in the M81 galaxy \citep{Bhardwaj2021_1,Kirsten2022} which does not seem to have any recent star formation. Using a sample of 37 localized FRBs and their spatial distribution with respect to the host galaxies, the authors of \cite{Gordon2025} showed that the majority of FRB progenitors are associated with massive stars with a minority formed in dynamic channels such as globular clusters which constitute $11\pm 5$ percent of the total population. However, there was no definitive proof of high star formation driving efficient FRB formation.

Instead of focusing on individual FRBs, several works \citep{ZZLL2021,James2022_1} have tried to study the whole population of FRBs with available data. This has the advantage that our inference is sensitive to the properties of the average population of FRBs and is not driven by a few outlier events. Using a catalogue of FRBs detected by the CHIME collaboration \citep{Chime_catalog}, a recent work \citep{Gupta2025} constrained the progenitor model of FRBs. They find that a model where the FRB population depends upon a combination of star formation rate and stellar mass of host galaxy, describes the present data very well. Another work, \cite{HM2025}, used a sample of $\sim 50$ localized FRBs, with measured star formation rate and host galaxy stellar mass, to arrive at a similar conclusion. 

Typically, in the FRB population analysis, a template of cosmological star formation rate is used \citep{Gupta2025}. This template is derived by fitting directly measured star formation rate data from surveys at $z\lesssim 4$ \citep{MD2014}. In doing so, it is implicitly assumed that all galaxies contribute to FRB formation. However, it is possible that only a subset of all galaxies are likely to potentially host FRBs. This may affect our derived conclusions from such an analysis regarding the efficiency of FRB formation in these galaxies. In this paper, we use the observed galaxy stellar mass function to compute the cosmological star formation rate and stellar mass density. This construction can naturally capture any preference for FRB hosting galaxies. \color{black} As a proof of concept, \color{black} using the empirically observed distribution for stellar mass of host galaxies \color{black} at face value, \color{black} we find that the relevant cosmological stellar mass density and star formation rate may have been over-estimated in previous works. We also find that bias can be introduced in inferring parameters if we do not take into account this distribution, given a population of observed FRBs.   

The ideas presented in this paper are not specific to FRBs as such and are applicable to the distributions of, e.g., Gamma Ray Bursts \citep{PGSGVNSMCG2016,WP2010,PD2021} and Gravitational wave (GW) events \citep{LIGO2023}. Unlike FRBs, it is difficult to localize the host galaxy of merging binary black holes (BBHs), using GW data alone. The cosmological distribution of merging BBHs provide an indirect way to probe the host galaxy properties of these objects. In the data analysis step, as in the case of FRBs, a similar template of cosmological star formation is assumed in the case of BBHs. Using such a template, the authors in \cite{Vijaykumar2024} showed that the distribution of BBH is consistent with host galaxies weighted by their star formation rate and less than about 43 percent (90 percent confidence) of mergers follow a sample of  stellar mass weighted galaxies. In this work, we show that this interpretation may be biased. Alternatively, we can use the galaxy stellar mass function in conjuction with cosmological distribution of BBHs to infer the host galaxy properties more accurately.

This paper is structured as follows. We give a brief description of previous works regarding cosmological distribution of FRBs in Sec. \ref{sec:cosmological_FRB}. We introduce the galaxy stellar mass function and the relevant observations in Sec. \ref{sec:galaxy_SMF}. In Sec. \ref{sec:SMD_FRB}, we provide our main results and show how it differs from previous analysis. We test the hypothesis that FRB sources fully track the stellar mass density in Sec. \ref{sec:mass_weighting}. In Sec. \ref{sec:fY_inference}, we show the importance of our approach for inference of  parameters for cosmological distributions of FRBs. We discuss the potential impact of redshift evolution of FRB host galaxies' stellar mass in Sec. \ref{sec:redshift_evolution}. The application to the analysis of GW events is given in Sec. \ref{sec:GW_data}. We end with conclusions in Sec. \ref{sec:conclusions}.

%----------------------------------------------
\section{Cosmological FRB distribution with global star formation rate and stellar mass}
\label{sec:cosmological_FRB}
%----------------------------------------------
We use the theoretical setup of \cite{Beniamini2021}, \cite{Gupta2025} to relate the distribution of FRBs to cosmological observables. The FRBs are expected to mostly originate from highly active star-forming regions. The birth rate of magnetars can be related to the production rate of neutron stars \citep{Beniamini2021} which is related to star formation rate (SFR). Therefore, as a starting point, the FRB formation rate per volume is assumed to be proportional to the star formation rate per unit comoving volume or star formation rate density (SFRD) which is a function of redshift $z$. A fit was obtained by \cite{MD2014} after converting observed UV and infrared luminosity from galaxy surveys at $z<4$ to cosmological SFRD. We refer to this fit as Madau-Dickinson or "MD-fit" in this work. Recent observations with James Webb Space Telescope (JWST) show that the extrapolated MD-fit overestimates the SFRD at higher redshift. Therefore, we use an "improved MD-fit" at $z>4$ which was derived in \cite{Gupta2025}. Altogether, the expression for SFRD is given by,
%--------------------------------------------
    \begin{align} \label{eq:SFRD}
    \dot{m}_*(z)=0.015\frac{(1+z)^{2.7}}{1+[(1+z)/2.9]^{5.6}}\hspace{0.1cm} {\rm M_{\odot}yr^{-1}Mpc^{-3}},\hspace{0.2cm} z<4
    \nonumber \\
    \dot{m}_*(z)=10^{-0.257z-0.275}\hspace{0.1cm} {\rm M_{\odot}yr^{-1}Mpc^{-3}},\hspace{0.8cm} 4<z<10
\end{align}
%--------------------------------------------
However, recent evidence suggest that the assumption of SFR being the sole driver of FRB formation may be wrong \cite{Gupta2025,HM2025}. These works suggest that the stellar mass of the host galaxy also plays a role in FRB formation and the current available data is best explained by a model where FRB formation tracks a mixture of SFR and stellar mass. The authors of \cite{Gupta2025} used a sample of unlocalized FRBs from the CHIME FRB collaboration \cite{Chime_catalog}. Since the redshifts are unknown, they used dispersion measures (DM) and observed fluxes of FRBs as indirect measures to constrain the redshift distribution of this sample of FRBs. They fit the inferred redshift distribution of FRBs as a linear combination of populations tracking SFRD and stellar mass density (SMD) or $m_*$. The expression for $m_*$ is given by,
%--------------------------------------------
\begin{equation}
    m_*(z)=(1-R)\int_z^{\infty}\frac{\dot{m}_*(z')}{(1+z')H(z')}{\rm d}z', 
    \label{eq:SMD}
\end{equation}
%--------------------------------------------
where $R=0.27$ is the return fraction for Salpeter IMF. Therefore, SMD is just the time integral of SFRD with an overall correction factor. The authors conclude that possibly $31_{ -21}^{+31}\%$ of FRB sources track star formation or SFRD.  

 The expression for the FRB volumetric rate function is given by the expression \citep{Gupta2025},
%------------------------------------------
\begin{equation}
 %   R(z)=\frac{{\rm d}N}{{\rm d}z}=\Phi_0\psi_*(z)\frac{4\pi d_{\rm com}^2(z)}{(1+z)}\frac{c}{H(z)}\int P(E_{\nu}){\rm d}E_{\nu}
     R(z)=\frac{{\rm d}N}{{\rm d}V{\rm d}t}=\Phi_0\psi_*(z)\int P(E_{\nu}){\rm d}E_{\nu}
    \label{eq:FRB_rate_gupta}
\end{equation}
%------------------------------------------
where $N$ is the number of observed FRBs, $\Phi_0$ is the observed volumetric rate of FRBs at $z=0$, 
%$d_{\rm com}(z)$ is the comoving distance at $z$ 
and $P(E_{\nu})$ is the probability density function for the specific energy $E_{\nu}$ in the comoving frame of FRB. The intrinsic specific energy distribution of FRBs has a frequency dependence with, $E_{\nu}=E_{{\nu}_{\rm com}}\left(\frac{\nu}{\nu_{\rm com}}\right)^{\alpha}$ with $\nu_{\rm com}=\nu_0(1+z)$ with $\nu_0=600$ MHz. 

The observed FRB rate function is related to SFRD and SMD through the factor $\psi_*(z,f_Y)$. The expression for $\psi_*(z,f_Y)$ is given as,
%-----------------------------------------
\begin{equation}
    \psi_*(z,f_Y)=f_Y\frac{\dot{m_*}(z)}{\dot{m_*}(z=0)}+(1-f_Y)\frac{m_*(z)}{m_*(z=0)},
    \label{eq:psi_z_eq}
\end{equation}
%-----------------------------------------
where $f_Y$ is the fraction of FRB sources tracking the SFRD while the rest track SMD. We use $f_Y=0.3$ as the fiducial value in the rest of the work which is roughly the best fit value obtained in \cite{Gupta2025}. We note that $f_Y$ was assumed to be redshift independent but, in general, it can have redshift dependence. We can rewrite Eq. \ref{eq:FRB_rate_gupta} as,
%------------------------------------------
\begin{equation}
    R(z)=\left(\frac{\Phi_0}{\dot{m_*}(z=0)}\right)\dot{m_*}(z)f_Y+\left(\frac{\Phi_0}{{m_*}(z=0)}\right){m_*}(z)(1-f_Y),
    \label{eq:R_z}
\end{equation}
%------------------------------------------
where we have suppressed the factors of $\int P(E_{\nu}){\rm d}E_{\nu}$ in Eq. \ref{eq:FRB_rate_gupta} for brevity which will not affect the point that we are trying to make here. The terms in the bracket can be thought of as efficiency factors or rate of FRB formation per stellar mass or star formation rate. This formulation reproduces the observed FRB volumetric rate at $z=0$. 

%We note that the normalization of $m_*$ and $\dot{m}_*$ at $z=0$ has been absorbed into $\Phi_0$. Therefore, the prediction for $\Phi_0$ depends upon the value of $m_*(z=0)$ and $\dot{m_*}(z=0)$. 
%The best fit value of $f_Y\lesssim 1$ also shows that the global SMD or SFRD roughly reproduces the observed volumetric rate of FRB sources.

We note that the authors in \cite{Gupta2025} assumed SFRD and SMD to be functions of redshift alone (which we refer to as ``global" quantities). They have used expression for $\dot{m_*}(z)$ and ${m_*}(z)$ from Eq. \ref{eq:SFRD}.  By definition, these quantities  account for star formation in all observed galaxies in the universe. However, it is possible that only a specific sub-population of galaxies can potentially host FRBs, for example, the host galaxies may have a preferred mass with some spread around that value. We compile a list of 53 localized FRBs with their measured stellar mass which we show in Table \ref{tab:FRB_sample}. Their distribution is plotted in Fig. \ref{fig:stellarmass_dist}. 
%We clearly see that galaxies with stellar mass $\sim 10^{10}$ $M_{\odot}$ dominate as potential FRB host galaxies with both higher and lower masses being less favoured potential hosts. 
At present, on average, the observed FRB host galaxies have a stellar mass of the order of $\sim 10^{10}$ $M_{\odot}$. \color{black} However, this may be manifestation of some potential selection effects. In this work, we use this distribution at face value to showcase the importance of galaxy stellar mass function in FRB population analysis. \color{black} We note that these hosts are located at $z\lesssim 0.5$. In the rest of the work, we assume that the host galaxies at higher redshifts have the same typical mass unless stated otherwise.  If we only consider this subsample of potential FRB host galaxies, the corresponding $m_*$ and ${\dot m_*}$ in Eq. \ref{eq:R_z} can be significantly different. This would suggest a modified efficiency factor of FRB formation per stellar mass or star formation rate as compared to the fiducial case in order to reproduce the observed FRB rate at $z=0$. To study this in more detail, we first study the abundance of galaxies as a function of stellar mass and redshifts, in the next section. 

%---------------------------------------------------
\section{Galaxy Stellar Mass Function (SMF)}
\label{sec:galaxy_SMF}
%---------------------------------------------------
The galaxy SMF quantifies the number density of galaxies per comoving volume as a function of stellar mass and redshift. In this work, we use the SMF obtained using Cosmic Dawn Survey Pre-launch catalogue \citep{Euclid2025}. The survey covers an area of roughly 10.13 deg$^2$ and provides a catalogue of galaxies up to $z\sim 6$ in 11 redshift bins. Due to an order of magnitude increase in survey volume, this sample has lower cosmic variance than any other previous surveys. The catalogue has a galaxy stellar mass limit (lower mass limit of observed galaxies) of $10^9$ $M_{\odot}$ at $z\lesssim 1$ which increases to $10^{10}$ $M_{\odot}$ at $z\sim 6$. The authors fit a double and single Schechter function to the observed SMF at $z\lesssim 2$ and $z\gtrsim 2$ respectively. The double Schechter function can be written as,
%-----------------------------------------------
\begin{equation}
    \Phi\mathcal{(M)}{\rm d}\mathcal{M}=\left[\Phi_1\left(\frac{\mathcal {M}}{\mathcal M_*}\right)^{\alpha_1}+\Phi_2\left(\frac{\mathcal {M}}{\mathcal M_*}\right)^{\alpha_2} \right]{\rm exp}\left(-\frac{\mathcal M}{\mathcal M_*}\right)\frac{{\rm d}{\mathcal M}}{\mathcal M_*}
    \label{eq:schechter}
\end{equation}
%----------------------------------------------
where the fitting parameters for redshift bins are provided in Sec. \ref{app:SMF_fit}. The variable $\mathcal{M}$ is equivalent to the stellar mass $M_*$. In Eq. \ref{eq:schechter}, we have kept the notation used in \cite{Euclid2025} for better readability and comparison. Using the Schechter fit, the authors extend the SMF to $10^8$ $M_{\odot}$. 
In this work, Eq. \ref{eq:schechter} is integrated
over stellar masses in the range $10^8-10^{13}$ $M_{\odot}$ to obtain SMD at a given redshift. In order to obtain a smooth, continuous distribution over redshift, we do least square fitting of the parameters in Schechter function. More details are provided in Sec. \ref{app:SMF_fit}. 

In a complementary fashion, using deeper JWST data, COSMOS-Web survey \citep{COSMOSWEB2025} provides galaxy SMF up to $z\sim 12$. In this case, the galaxy stellar limit is $10^8$ $M_{\odot}$ which increases to $10^9$ $M_{\odot}$ at $z\sim 10$. However, the catalogue is cosmic variance dominated due to the smaller observed volume (0.53 deg$^2$) especially at $M_*\gtrsim 10^{10}$ $M_{\odot}$. This results in high stochastic fluctuation in the SMF at relatively higher stellar masses which becomes more severe at higher redshifts. In order to avoid these issues, we use the Cosmic Dawn Survey fits to infer our results. We matched the extrapolated results from Cosmic Dawn Survey data with COSMOS-Web at $M_*\lesssim 10^9$ $M_*$ and find similar results for a given redshift bin. With more data that will be gathered over time, we will have a more complete galaxy SMF survey over a larger range of stellar masses.  

%------------------------------------------------
\section{SMD from potential FRB hosting galaxies}
\label{sec:SMD_FRB}
%------------------------------------------------

%-----------------------------------------------
\begin{figure}[!htp]
\begin{subfigure}[b]{0.4\textwidth}
\includegraphics[scale=0.3]{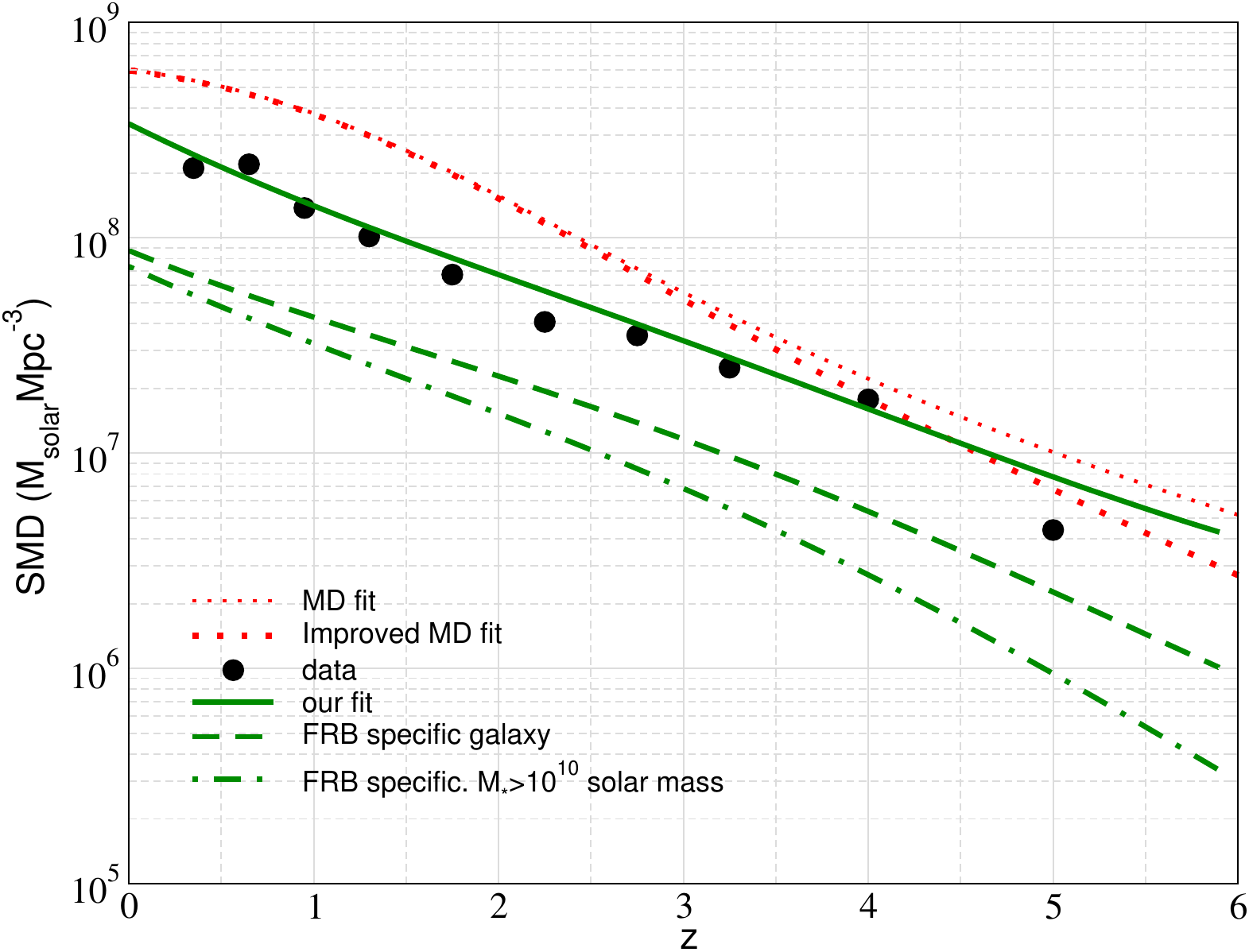}
%\caption{70 GHz}
%\label{fig:depfracz=1000}
\end{subfigure}\hspace{50 pt}
\begin{subfigure}[b]{0.4\textwidth}
\includegraphics[scale=0.3]{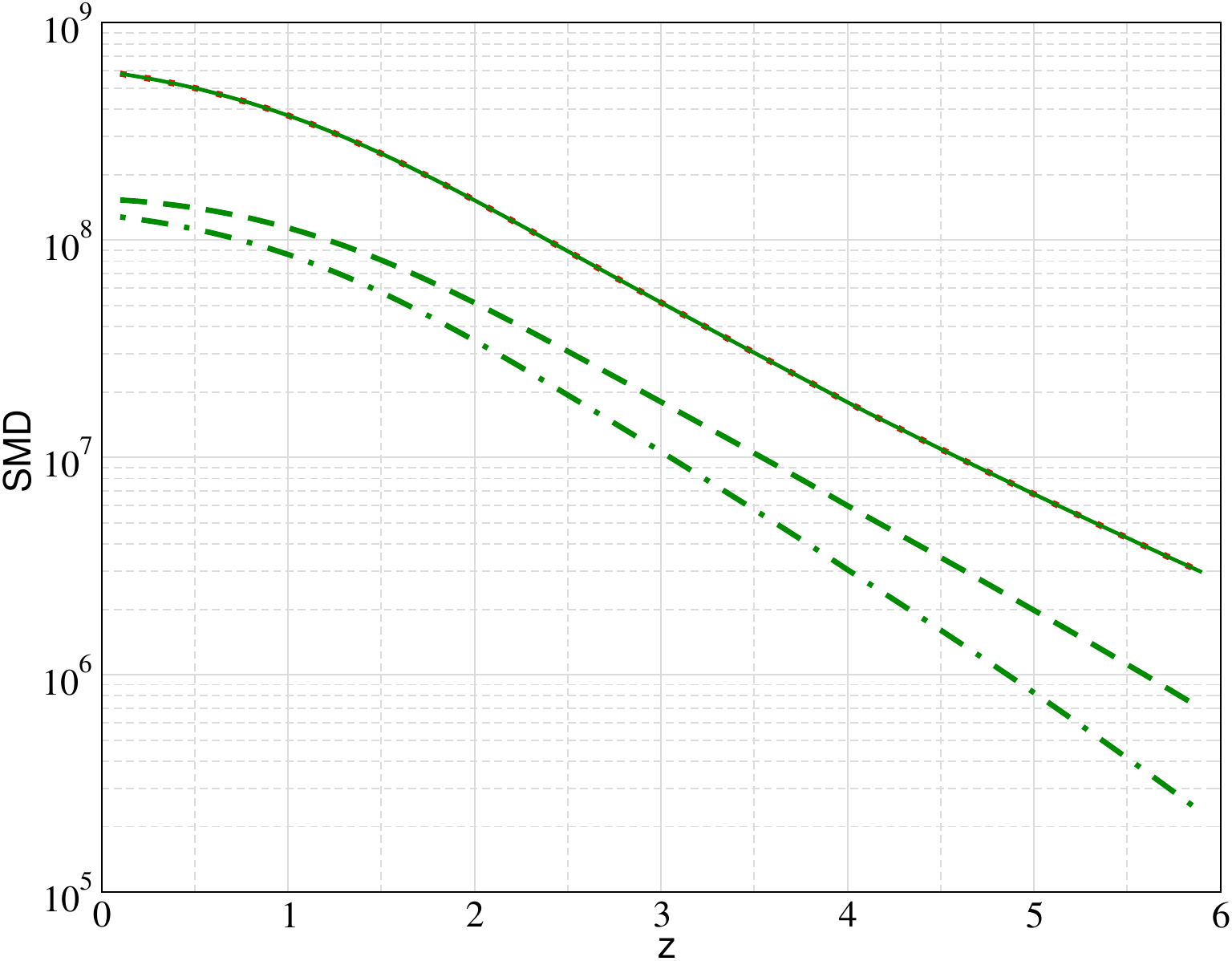}
%\caption{100 GHz}
%\label{fig:depfracz=100}
\end{subfigure}
\caption{Evolution of SMD as a function of redshift. (Left panel) The inferred SMD from galaxy SMF is shown in black points. Our smooth fit to the data points is shown in solid green. We consider galaxies with masses between $10^8-10^{13}$ $M_{\odot}$. In dashed green, we plot the SMD from potential FRB host galaxies with an appropriate filter function. We also plot a scenario where we isolate the contribution from host galaxies with $M_*> 10^{10}$ $M_{\odot}$. (Right panel) We scaled the SMD inferred from galaxy SMF, in a redshift dependent way, to match the improved MD fit. We apply the same scale factor to the other curves in green in the left panel. }
 \label{fig:euclid_data_fit}
\end{figure}
%--------------------------------------------------

The expression of SMD is given by,
%------------------------------------------
\begin{equation}
    {\rm SMD}(z)=\int_{10^8 M_{\odot}}^{10^{13}M_{\odot}}\Phi(\mathcal M,z)F(\mathcal{M},z)\mathcal{M} {\rm d}\mathcal{M}, 
    \label{eq:SMD_SMF}
\end{equation}
%-----------------------------------------
where $F(\mathcal{M},z)$ \color{black} can be selection or \color{black} a filter function which is 1 if all galaxies are potential FRB hosts and FRBs are distributed in galaxies weighted by their stellar mass. \color{black}In general, the filter function can be a complicated function of stellar mass, star formation rate, metallicity of galaxies etc. which one needs to jointly fit along with templates of SFRD and SMD to a sample of localized FRBs. A joint fitting of all parameters along with filter function may not be possible at present as we need objects at high redshifts which is useful in breaking degeneracies between parameters since our SFRD and SMD templates have significant redshift evolution inherited from galaxy stellar mass function. In this paper, we use a very simplified approach with the filter function being just a function of stellar mass. \color{black}

%As we show in Sec. \ref{app:stellarmass_data}, the present observed data seems to prefer galaxy mass from an approximate lognormal distribution with mean $\log_{10}(\mathcal{M})\approx 10 $. %The function $F(\mathcal{M})$ captures this information. 
We plot the SMD obtained from galaxy SMF in Fig. \ref{fig:euclid_data_fit}.  We note that the SMD obtained from galaxy SMF has a mismatch of factor of about 2 at $z\lesssim 2$ with the MD fit. This is in line with the past works \citep{HB2006,WTH2008} as well as in \cite{COSMOSWEB2025}. The difference can be attributed to systematics such as overestimated dust attenuation, uncertain UV luminosity to SFR
conversion factor etc. We show the SMD obtained for potential FRB host galaxies in dashed green curve. As we show in Sec. \ref{app:stellarmass_data}, the present observed data seems to prefer galaxy mass from an approximate lognormal distribution with mean $\log_{10}(\mathcal{M})\approx 10 $ and $\sigma\approx 0.6$. In order to distinguish the case with FRB specific case from the total SMD, i.e. integrated over all galaxies, we denote the former as ${\rm SMD_1}$ and latter as ${\rm SMD_0}$. We note that the ${\rm SMD_1} \approx \frac{1}{3}{\rm SMD_0}$ . We also show the case where we consider FRB host galaxies with stellar mass $M_*\gtrsim 10^{10}$ $M_{\odot}$. This is motivated by the practical issue of difficulty in detectability of lower mass galaxies as we go to higher redshifts. We find that the relevant SMD goes down faster at higher $z$. This is expected since heavier galaxies are rarer at higher redshifts. In order to better compare our results with the work of \cite{Gupta2025}, we  scale our ${\rm SMD_0}$ (in a redshift dependent way) in order to match the improved MD fit. The other cases are also scaled appropriately. This is shown in the right panel of Fig. \ref{fig:euclid_data_fit}. 

\begin{figure}
 \centering
 \includegraphics[width=0.6\columnwidth]{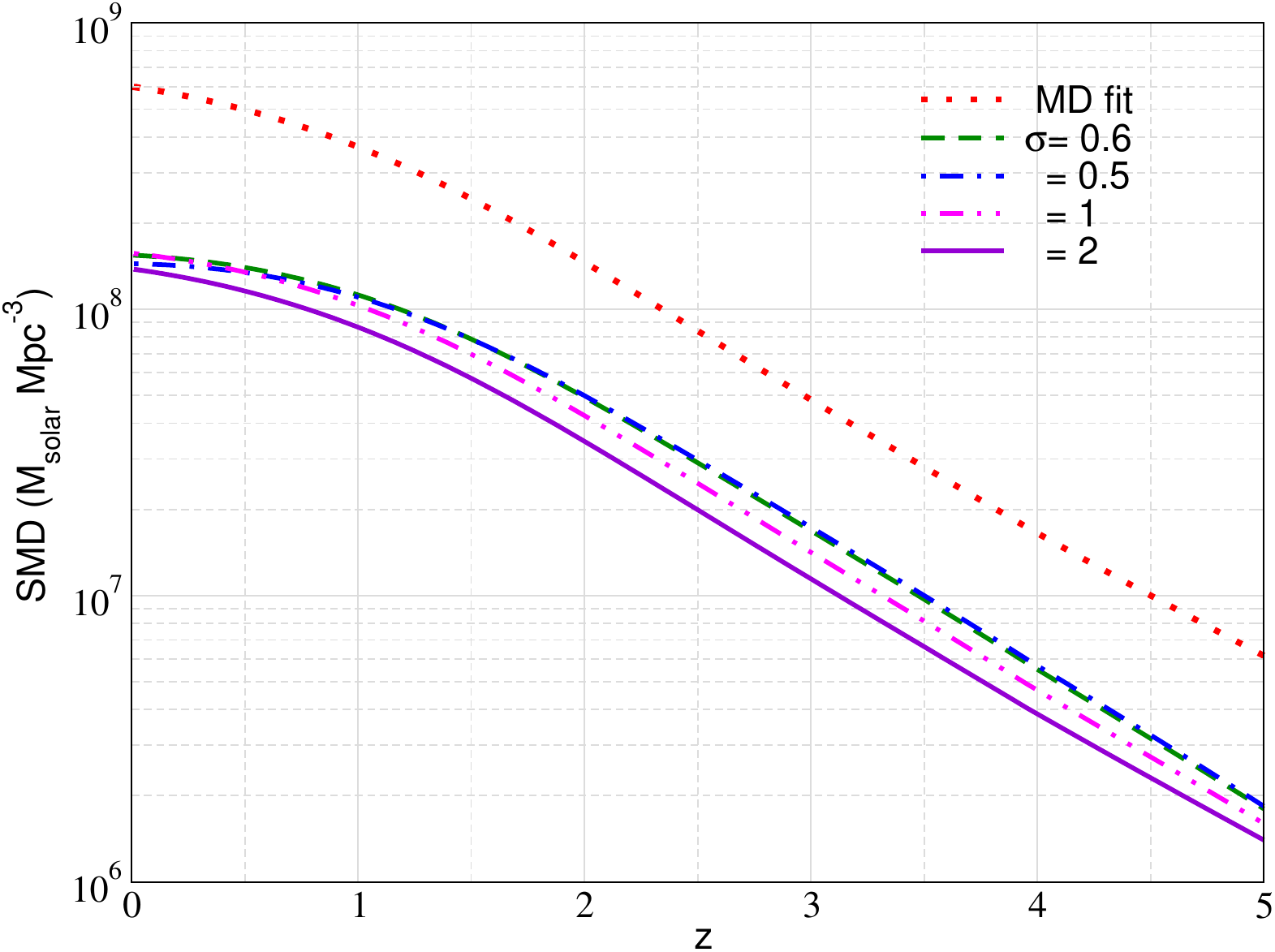}
 \caption{SMD from potential FRB host galaxies with varying $\sigma$ but same mean $\log_{10}(\mathcal{M})=10$ in the assumed filter function $F(\mathcal{M},z)$. }
 \label{fig:SMD_vs_sigma}
\end{figure}

As in Eq. \ref{eq:R_z}, we find that in order to reproduce the observed volumetric rate of FRBs at $z=0$, the FRB host galaxies have to be more efficient in producing FRBs than previously assumed. The FRB formation rate per stellar mass is given as $\frac{\Phi_0}{m_*(z=0)}$ which can be re-written as, $\left(\frac{\Phi_0}{{m_*}_{,\rm fid}(z=0)}\right)\left(\frac{{m_*}_{,\rm fid}(z=0)}{m_*(z=0)}\right)\approx 3\left(\frac{\Phi_0}{{m_*}_{,\rm fid}(z=0)}\right)$. Therefore, our results suggest that the FRB formation rate per stellar mass is boosted by a factor of 3 as compared to what was previously thought. Using the inferred volumetric rate of $\Phi_0=5.4\times 10^4$ Gpc$^{-3}$yr$^{-1}$ (fluence threshold of 1 Jy ms) \citep{Gupta2025,Shin2023}, our inferred $\frac{\Phi_0}{m_*(z=0)}$ turns out to be $2.7\times 10^{-13}$ $M_{\odot}^{-1}$yr$^{-1}$. Similarly, we find that the FRB formation rate per star formation rate is, $\left(\frac{\Phi_0}{\dot{m_*}_{,\rm fid}(z=0)}\right)\left(\frac{\dot{m_*}_{,\rm fid}(z=0)}{\dot{m_*}(z=0)}\right)\approx 2\left(\frac{\Phi_0}{\dot{m_*}_{,\rm fid}(z=0)}\right)$ which turns out to be $7.2\times 10^{-3}$ $M_{\odot}^{-1}$. \color{black} We clarify that these results are obtained assuming our choice of $F(M,z)$. In order to check the sensitivity of these results to choice of parameters of lognormal distribution,
we have used different shapes by varying the $\sigma$ of our lognormal filter function but with same mean $\log_{10}(\mathcal{M})=10$. The computed SMD by varying $\sigma$ is shown in Fig. \ref{fig:SMD_vs_sigma}. The changes due to varying $\sigma$ is less than a factor of 2 which can be attributed to the flatness of $\Phi(\mathcal{M})\mathcal{M}$ vs $\mathcal{M}$ at low redshifts. This computation assumes the mean value of lognormal distribution to be 10 which is anchored to observations at $z\lesssim 0.5$ which may not hold at higher redshifts. A more rigorous analysis would require extracting  filter function from data which is beyond the scope of this work.

\color{black}

%-------------------------------------------------
%------------------------------------------------

%This efficiency parameter can be computed by the ratio $\frac{{\rm SMD_0}}{\rm SMD_1}$ which we denote as $\epsilon(z)$ and is plotted in Fig. \ref{fig:efficiency_factor}. As noted earlier, $\epsilon(z)\approx 3-4$ at $z=0$ and is a function of redshift due to the dependence of galaxy SMF on $z$. 

%-------------------------------------------------
%\begin{figure}
% \centering
% \includegraphics[width=0.8\columnwidth]{./eps/fitting_scale.pdf}
% \caption{The efficiency parameter $\epsilon(z)$ (as defined in text) as a function of $z$. }
% \label{fig:efficiency_factor}
%\end{figure}
%------------------------------------------------

%------------------------------------------------
\subsection{Mass weighted galaxy SMF as potential FRB source}
\label{sec:mass_weighting}
\begin{figure}
 \centering
 \includegraphics[width=0.8\columnwidth]{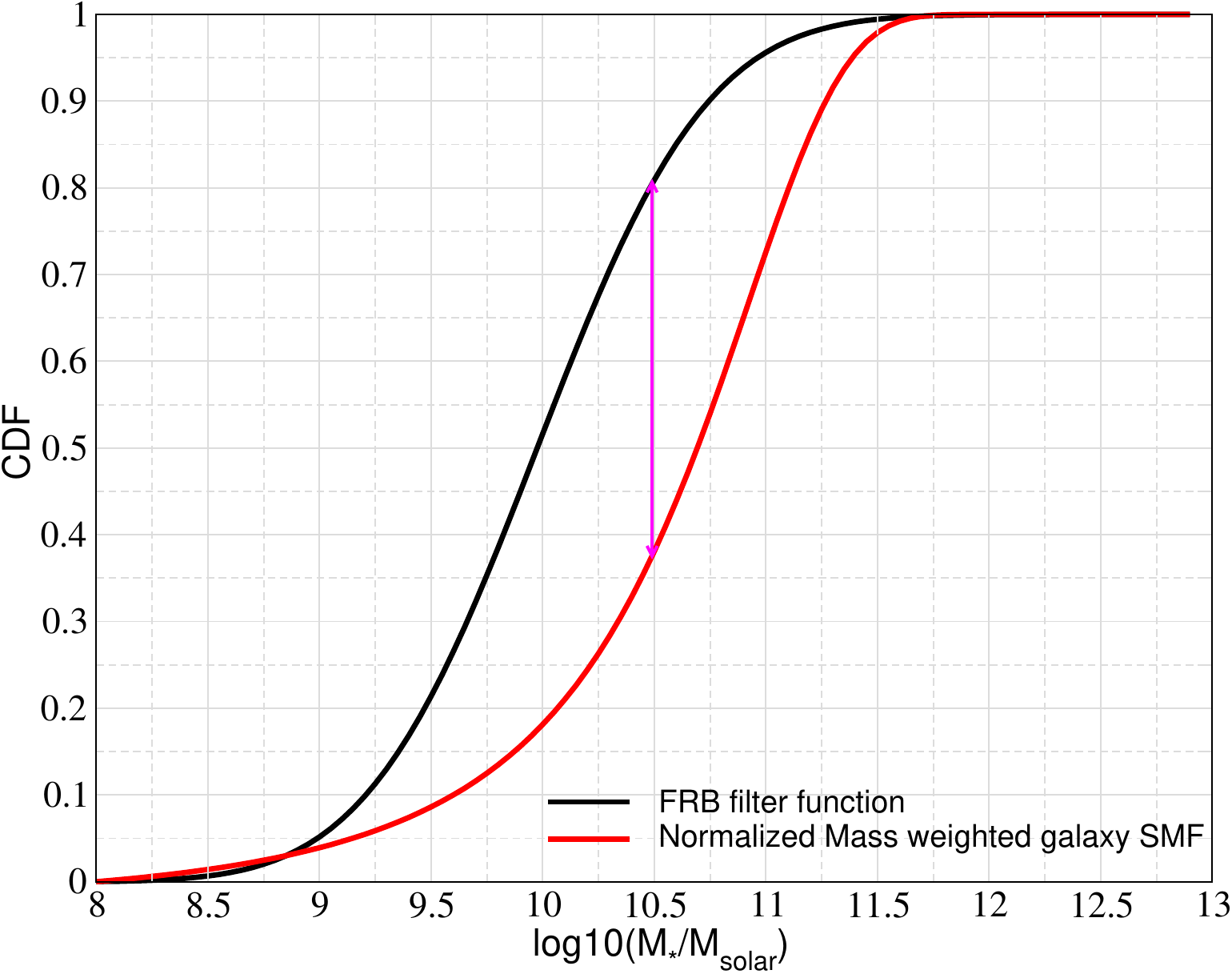}
 \caption{Cumulative distribution function (CDF) of FRB filter function and the normalized mass weighted galaxy SMF or $\Phi(\mathcal{M})\mathcal{M}$ at the lowest redshift bin with central $z=0.35$. The maximum difference between the two CDFs is shown in the double-headed magenta line. }
 \label{fig:CDF_FRB}
\end{figure}

%------------------------------------------------
As we have stated before, the evidence from current data suggests that FRB sources track a combination of SFR and stellar mass. Here, we test the hypothesis of FRB sources fully tracking the galaxy stellar mass but within the observed range of host stellar masses. We do this comparison in Fig. \ref{fig:CDF_FRB}. We plot the probability distribution function for potential FRB host galaxies in Fig. \ref{fig:stellarmass_dist} (Sec. \ref{app:stellarmass_data}). As has been already noted, this function peaks around $M_*\approx 10^{10}$ $M_{\odot}$. We compare it to the normalized mass weighted galaxy SMF distribution or $\Phi(\mathcal{M})\mathcal{M}$ at the lowest redshift bin with a central redshift of 0.35. The qualitative behaviour can be understood from Table \ref{tab:parameter_fit}. At lower mass, the distribution of $\Phi(\mathcal{M})$ is given by the shallower power law with index $\alpha_2$. Therefore, $\Phi(\mathcal{M})\mathcal{M}$ distribution is peaked at $\mathcal{M_*}$ beyond which it steeply declines. For the lower redshift bins, we see that $\mathcal{M_*}\approx 10^{11}$ $M_{\odot}$.

We do a Kolmogorov-Smirnov test to distinguish the two distributions. We find that the maximum difference between the two CDFs is $\approx 0.4$. For $\sim 50$ FRBs, the difference between the two distributions has to be less than 0.19 in order to be consistent at 95 percent confidence interval. Therefore, we conclude that FRB sources do not fully track SMD which is consistent with previous works.

%-------------------------------------------------
%------------------------------------------------

%----------------------------------------------
\subsection{Bias in inferring $f_Y$ with Madau fit}
\label{sec:fY_inference}
\begin{figure}[!htp]
\begin{subfigure}[b]{0.4\textwidth}
\includegraphics[scale=0.3]{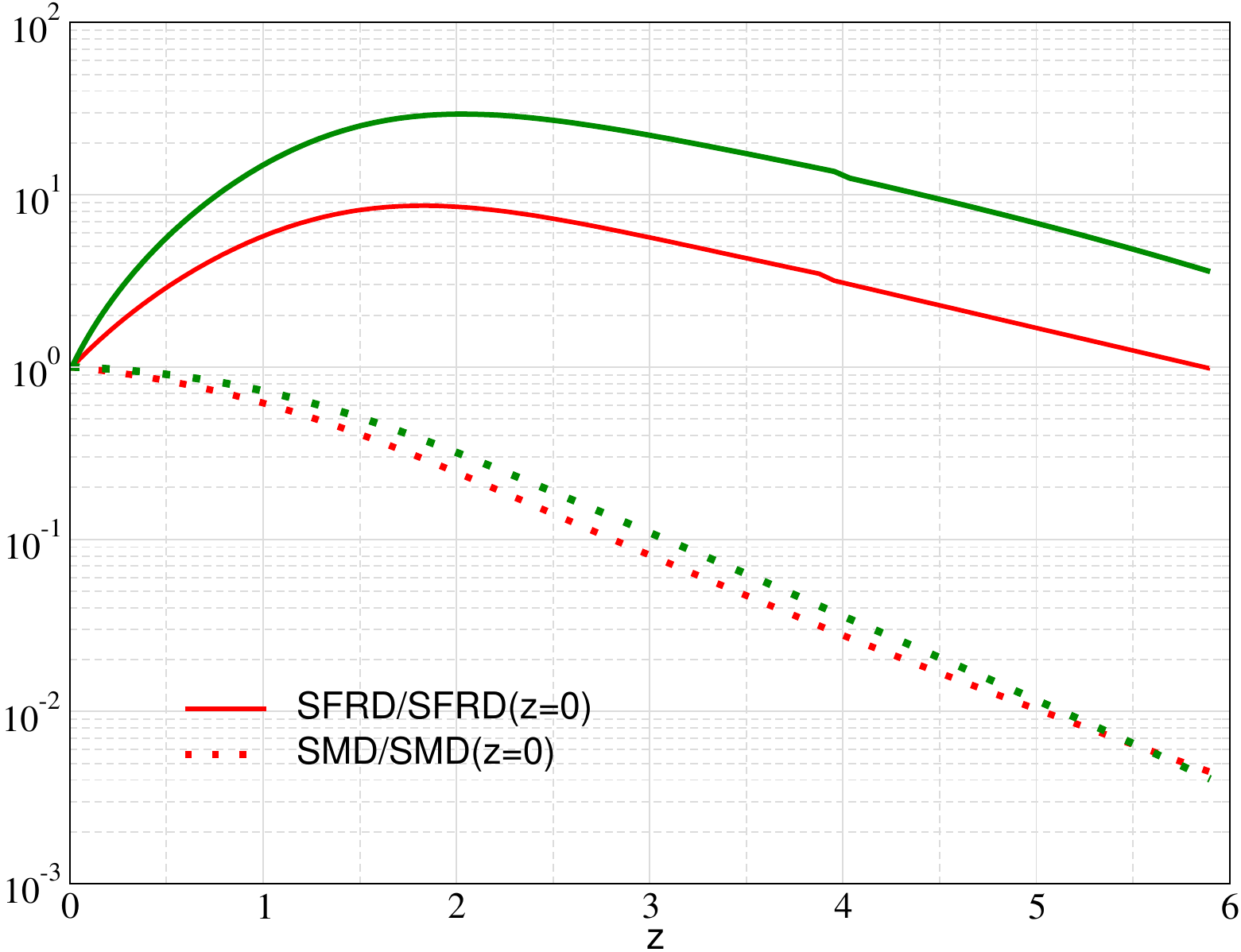}
%\caption{70 GHz}
%\label{fig:depfracz=1000}
\end{subfigure}\hspace{50 pt}
\begin{subfigure}[b]{0.4\textwidth}
\includegraphics[scale=0.3]{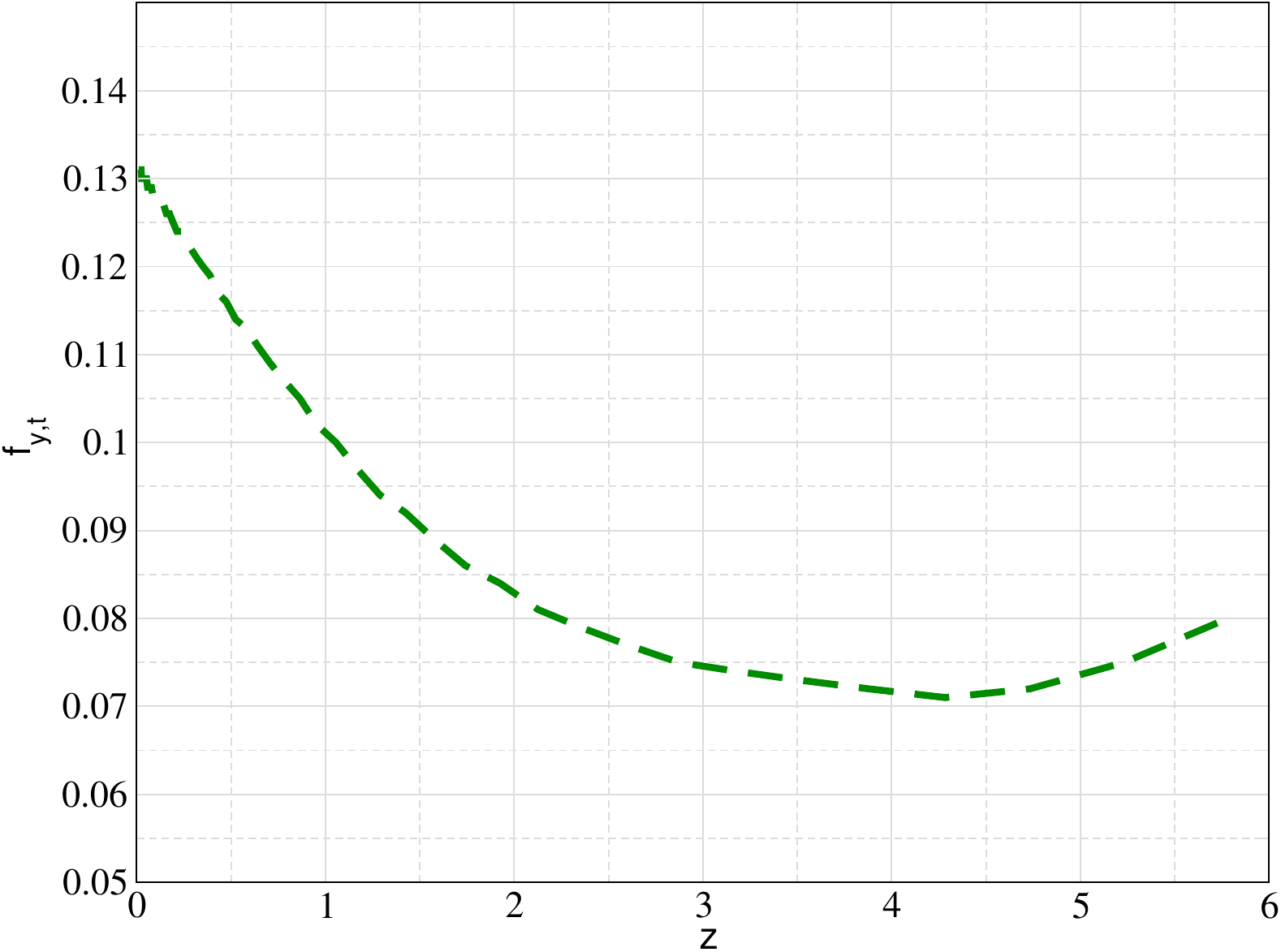}
%\caption{100 GHz}
%\label{fig:depfracz=100}
\end{subfigure}
\caption{Plot of ratio ${\rm SFRD/SFRD(z=0)}$ and ${\rm SMD/SMD(z=0)}$ for improved MD fit (red) and with galaxy SMF with FRB specific galaxies (green) (left panel). The inferred $f_{Y,{\rm t}}$ using the galaxy SMF where we  simulated FRB distribution with $f_Y=0.3$ (see text for details).    }
 \label{fig:psi_z}
\end{figure}

%----------------------------------------------
 For a cosmological distribution of FRBs correlated with galaxy SMF, we expect parameters such as $f_Y$ to be biased if inferred using the expressions of SFRD and SMD from the MD fit. From Eq. \ref{eq:FRB_rate_gupta}, we see that the redshift dependence of SMD and SFRD is captured in the function $\psi_*(z)$. We plot the individual components of $\psi_*(z)$ for the two cases in Fig. \ref{fig:psi_z}. We consider a hypothetical scenario where the cosmological distribution of FRBs is proportional to $\psi_*(z)$ obtained using the improved MD fit with $f_Y=0.3$. Then, we try to infer the corresponding $f_{Y,{\rm t}}$ using the SMF inferred functions. The observed volumetric rate of FRBs at $z=0$ is tuned to be same for the two cases. Therefore, the boost in efficiency factor at $z=0$ is captured in $\Phi_0$ and we are only interested in the redshift dependence of $f_{Y,{\rm t}}$ while $f_{Y,{\rm t}}$ is constrained to be less than 1.
 The inferred $f_{Y,{\rm t}}$ is plotted in the right panel of Fig. \ref{fig:psi_z}. We see that there is strong redshift dependence of inferred $f_{Y,{\rm t}}$. One can also do the reverse case, by generating a distribution with constant $f_{Y,{\rm t}}$ and show that the inferred $f_Y$ will be strongly redshift dependent. The reason of this dependence is the redshift evolution of galaxy SMF. We note that the quantitative results of this section strongly depends upon the galaxy SMF. As we have seen in Sec. \ref{sec:SMD_FRB}, there can be systematic issues regarding SMD measurements using SMF and the MD fit at $z\lesssim 2$. Therefore, the discussion in this section should be considered as more qualitative in nature and the quantitative details should not be taken at the face value.  

%-----------------------------------------------
%--------------------------------------------------

%--------------------------------------------------
\subsection{Implications of evolution of $F(\mathcal{M})$ with redshift}
\label{sec:redshift_evolution}
\begin{figure}
 \centering
 \includegraphics[width=0.8\columnwidth]{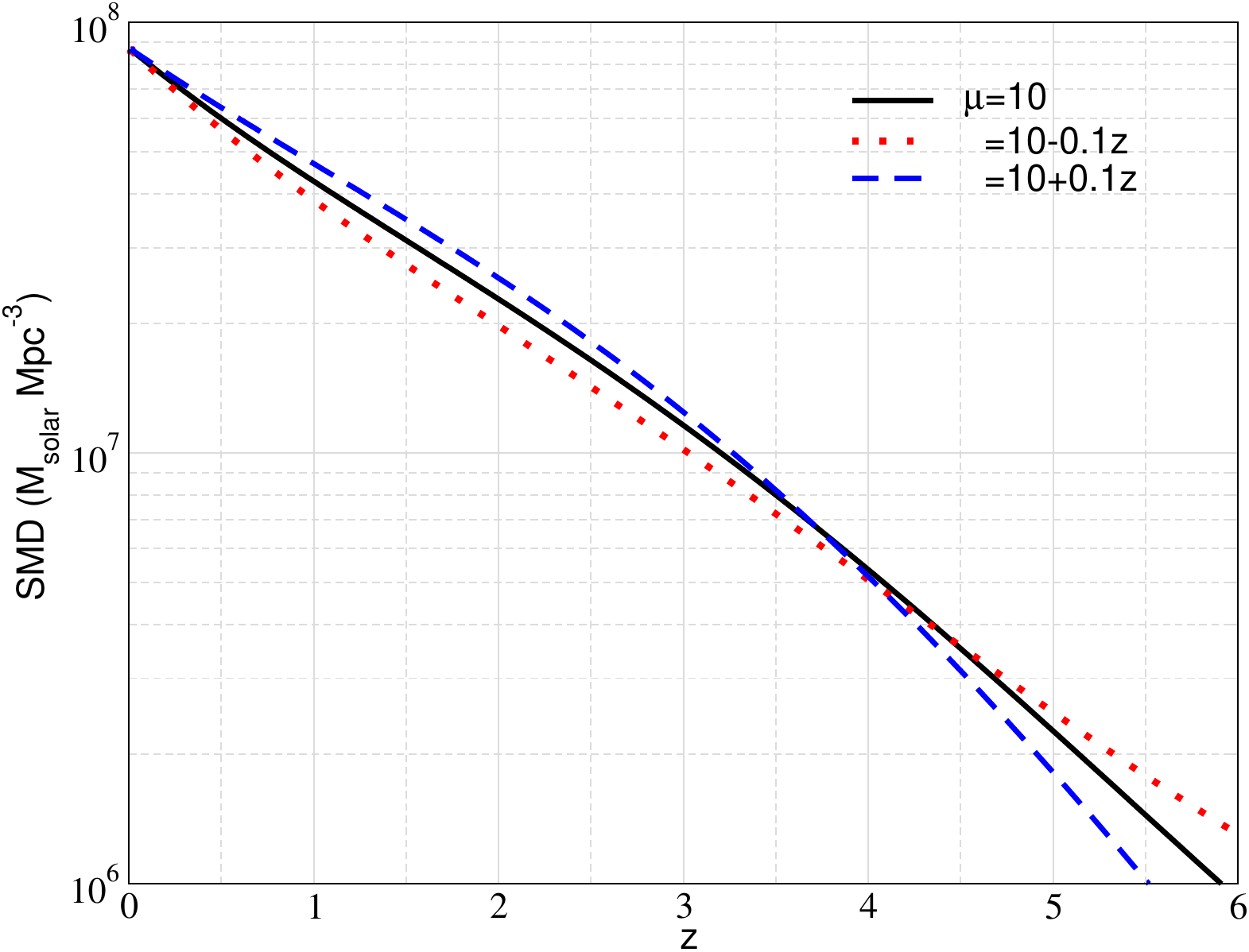}
 \caption{SMD using Eq. \ref{eq:SMD_SMF} with lognormal $F(\mathcal{M})$ and $\sigma=0.6$ and redshift dependent $\mu$ as shown in Fig. We have not applied any scaling to the SMD as opposed to the right panel of Fig. \ref{fig:euclid_data_fit}.  }
 \label{fig:SMD_mu}
\end{figure}

%--------------------------------------------------
Up to this point, we have assumed that potential FRB hosting galaxies or $F(\mathcal{M})$ in Eq. \ref{eq:SMD_SMF} do not have any redshift dependence.  This is due to limited present data with most of localized FRBs at $z\lesssim 0.5$. In this section, we study a few cases where the filter function $F(\mathcal{M})$ has some redshift dependence as shown in Fig. \ref{fig:SMD_mu}. We compute the quantity in Eq. \ref{eq:SMD_SMF} with a lognormal filter function with redshift-evolving mean $\mu$. We consider two cases where the mean stellar mass of FRB hosting galaxies ${\rm log10}(M_*)$ evolve linearly from 10 ($z=0$) to 11 or 9 ($z=10$). The two cases represent increasing or decreasing mean stellar mass of host galaxies. We do not see any significant changes with respect to the fiducial case. This may be due to our choice of mild redshift dependence. However, there is not significant indication of strong redshift dependence at present \citep{AB2025}. With more localized FRBs at $z\gtrsim 1$, we may be able test this assumption in future. We note that any redshift dependence can be naturally included using galaxy SMF which may not be possible by using just the MD fit. This also shows the relevance of our approach in this work.

%-------------------------------------------------
%------------------------------------------------

%----------------------------------------------
\section{Inferring host galaxy properties of binary black hole gravitational wave events}
\label{sec:GW_data}
\begin{figure}[!htp]
\begin{subfigure}[b]{0.4\textwidth}
\includegraphics[scale=0.3]{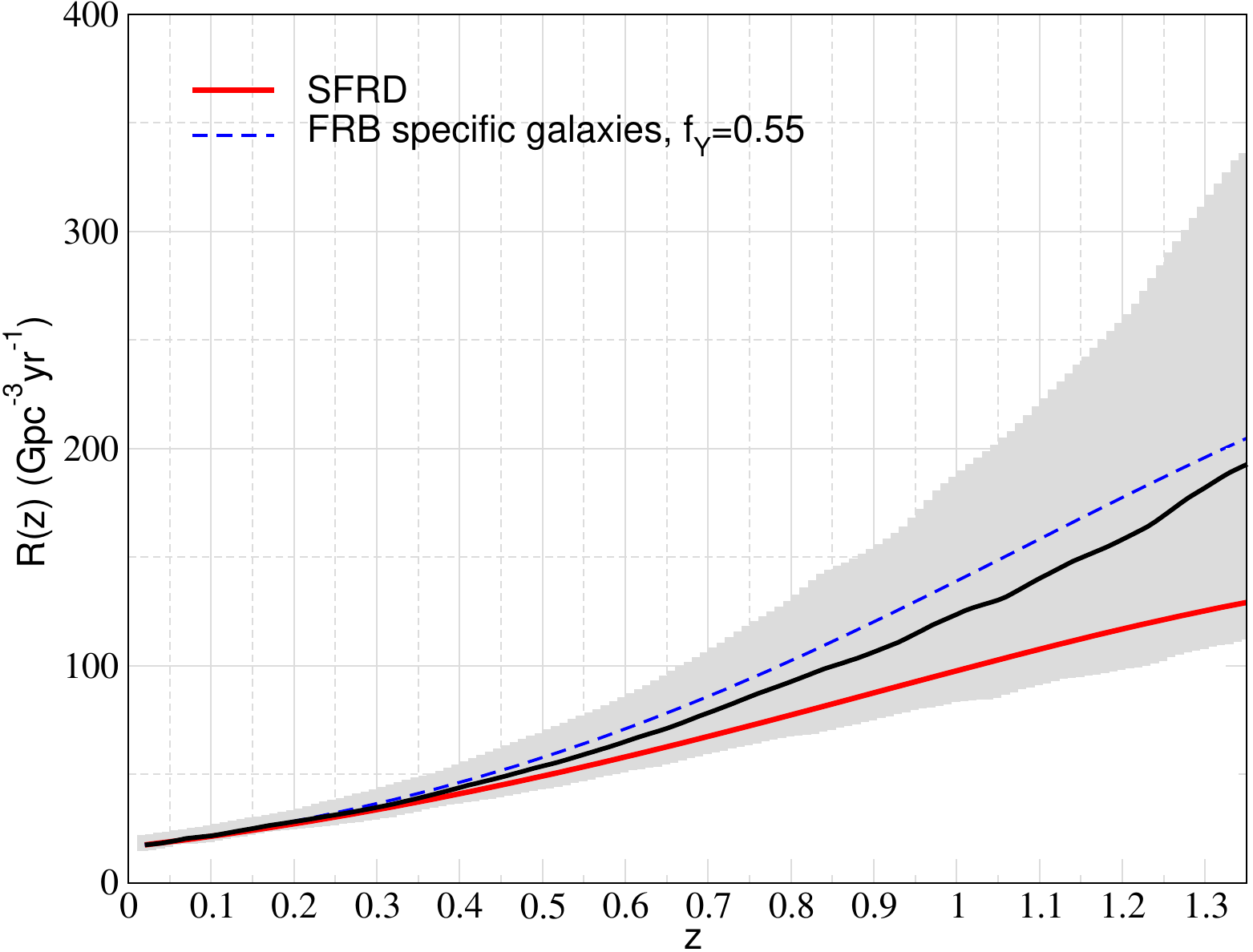}
%\caption{70 GHz}
%\label{fig:depfracz=1000}
\end{subfigure}\hspace{50 pt}
\begin{subfigure}[b]{0.4\textwidth}
\includegraphics[scale=0.3]{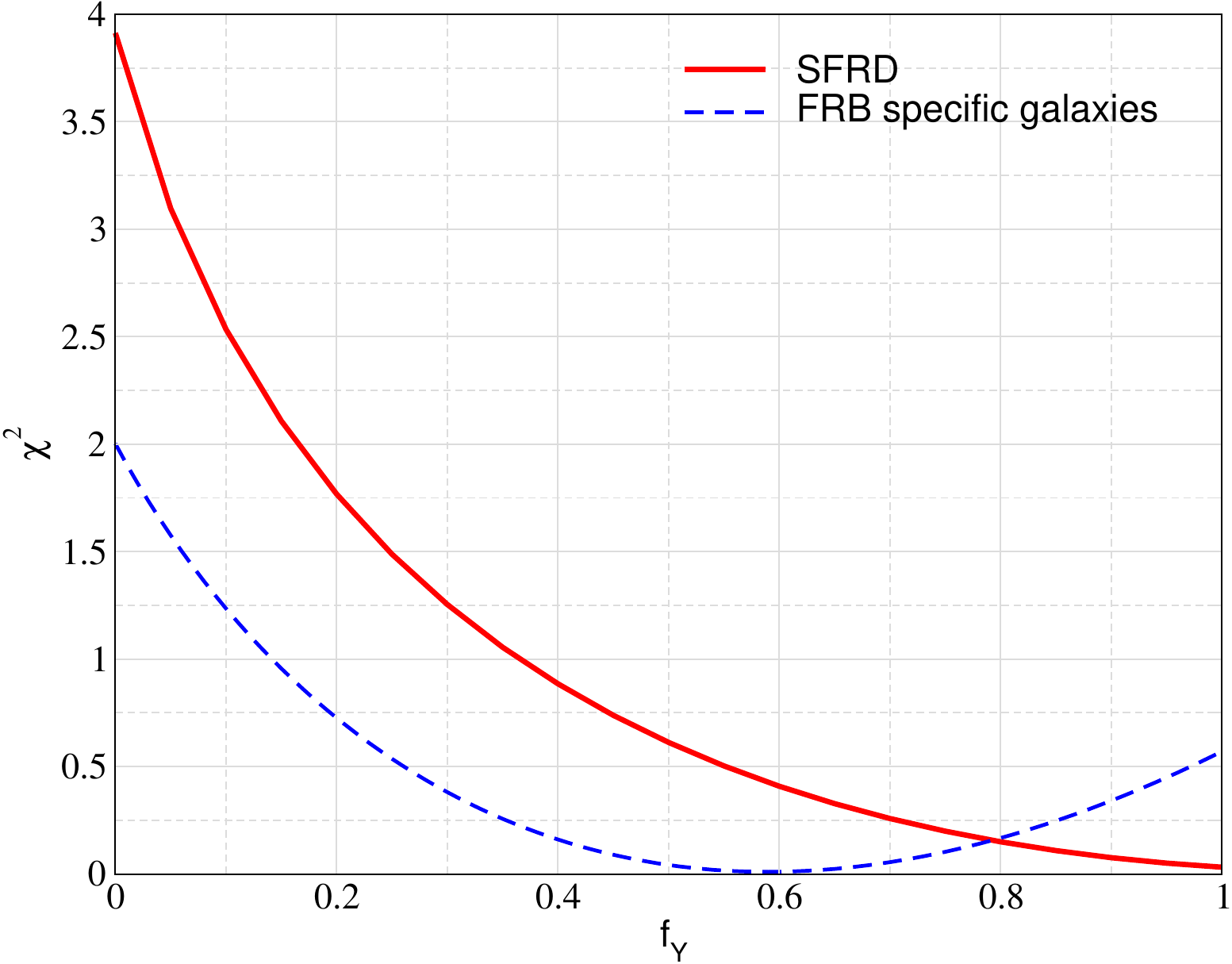}
%\caption{100 GHz}
%\label{fig:depfracz=100}
\end{subfigure}
\caption{The inferred merger rate of BBHs in the comoving frame as a function of redshift (left panel). The solid black line shows the best fit and the grey band show 50 percent confidence interval. The cosmological SFRD (Eq. \ref{eq:SFRD}) is shown in red line. The best fit assuming FRB specific galaxies (Sec. \ref{app:stellarmass_data}) is shown in blue dashed line. (Right panel) Value of $\chi^2$ as a function of $f_Y$ for two cases discussed in text. We caution the reader that the magnitude of $\chi^2$ should not be taken at face-value and we are only interested in the value of $f_Y$ where $\chi^2$ is minimum. }
\label{fig:GW_inference}
\end{figure}

%----------------------------------------------
While a fraction of FRBs have been localized to their host galaxies, the GW events from BBHs are unlikely to be localized in near-future experiments using only GW data. Therefore, the cosmological distribution of BBH mergers as a function of redshift provides an indirect way to infer their host galaxy properties. Using the GW event catalogue, the LIGO collaboration have inferred the volumtric merger rate of BBHs \citep{LIGO2023} in the comoving frame $R(z)$. This is shown in the left panel of 
Fig. \ref{fig:GW_inference}. Similar to FRBs, the expression for $R(z)$ can be written as,
%------------------------------------------
\begin{equation}
 %   R(z)=\frac{{\rm d}N}{{\rm d}z}=\Phi_0\psi_*(z)\frac{4\pi d_{\rm com}^2(z)}{(1+z)}\frac{c}{H(z)}\int P(E_{\nu}){\rm d}E_{\nu}
     R(z)=\frac{{\rm d}N}{{\rm d}V{\rm d}t}=\Phi_{\rm 0,BBH}\psi_*(z,f_Y)
    \label{eq:BBH_rate}
\end{equation}
%------------------------------------------
where $\Phi_{\rm 0,BBH}$ is the volumetric merger rate at $z=0$ and $\psi_*(z,f_Y)$ is still given by the expression in Eq. \ref{eq:psi_z_eq}. In Fig. \ref{fig:GW_inference}, we also compare two cases, (1) $\psi_*(z,f_Y)$ is given by SFRD and SMD from Eq. \ref{eq:SFRD}  and (2) $\psi_*(z,f_Y)$ for FRB specific host galaxies (Sec. \ref{app:stellarmass_data}). To obtain the best fit $f_Y$ value, we do a $\chi^2$ test where we fit the predicted rate of mergers today to the observed rate. The merger rate in observer reference frame is given by ,
%------------------------------------------
\begin{equation}
 %   \frac{{\rm d}N}{{\rm d}z}\vert=\Phi_0\psi_*(z)\frac{4\pi d_{\rm com}^2(z)}{(1+z)}\frac{c}{H(z)}\int P(E_{\nu}){\rm d}E_{
 \frac{{\rm d}N}{{\rm d}t{\rm d}z}=\Phi_{\rm 0,BBH}\psi_*(z)\frac{{\rm d}V}{{\rm d}z}\left(\frac{1}{1+z}\right)
    \label{eq:BBH_rate_today}
\end{equation}
%------------------------------------------
We divide the data points to 30 linearly-spaced redshift bins and sum over them to compute the $\chi^2$. We note that this binning is artificial and the value of $\chi^2$ will change with different choice of binning. However, we are only interested in the best fit value of $f_Y$ where $\chi^2$ is minimized. We have checked with different binning that the best fit $f_Y$ does not change. We plot the $\chi^2$ distribution as a function of $f_Y$ and we find that the best fit value turns out to be 1 for case (1) (see above) and 0.55 for the case (2). We plot these two cases in the left panel of Fig. \ref{fig:GW_inference}. We find that the two cases provide reasonable fits to the observed BBH rate at 1$\sigma$ level. We first compare our results with the conclusions of \cite{Vijaykumar2024}. The authors assumed the SFRD and SMD to be of the form in Eq. \ref{eq:SFRD} (our case 1) and compared with the data. The authors conclude that the data is consistent with SFRD while only less than 43 percent (90 percent confidence) of host galaxies weighted by their stellar mass can contribute to the observed BBH distribution. \color{black} In our setup, we obtain the same SFRD and SMD,  when we put the filter function $F(M,z)=1$. With this specific assumption, we obtain similar conclusion where our best fit $f_Y$ for the same case is 1.  Relaxing the above assumption and assuming FRB specific host galaxies, \color{black} we find that the observed BBH merger rate can still be explained by a 50:50 mixture of host galaxies weighted by their stellar mass and star formation rate respectively. 
%For this case, we have assumed that the observed BBH mergers are hosted in similar galaxies as FRBs. 
We acknowledge that there is no strong physical evidence for such a claim. The point is just to give an illustrative example. \color{black} Therefore, we conclude that allowing for different $F(M,z)$  broadens the allowed solution space given the current GW data. \color{black}

\begin{figure}[!htp]
\begin{subfigure}[b]{0.4\textwidth}
\includegraphics[scale=0.3]{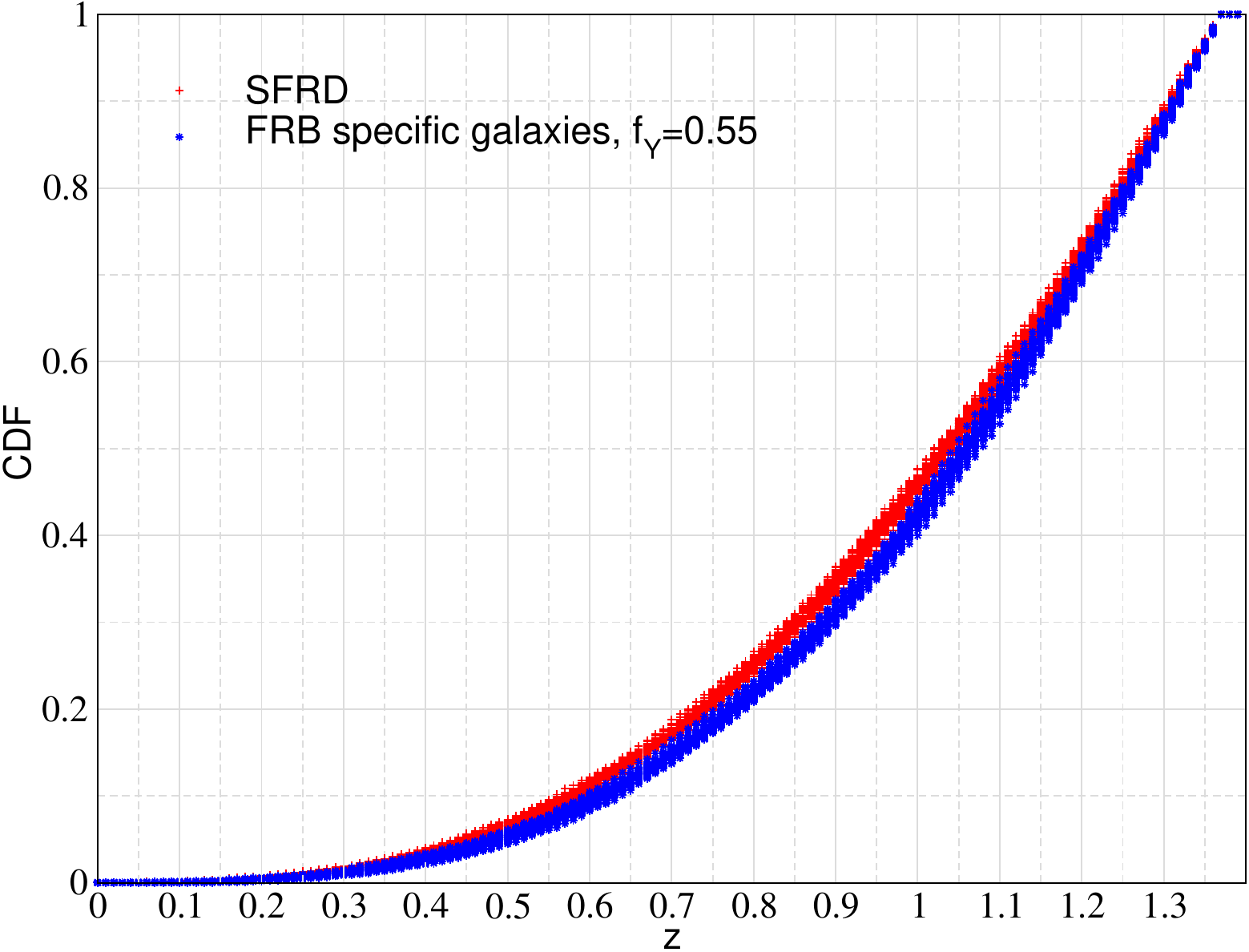}
%\caption{70 GHz}
%\label{fig:depfracz=1000}
\end{subfigure}\hspace{50 pt}
\begin{subfigure}[b]{0.4\textwidth}
\includegraphics[scale=0.3]{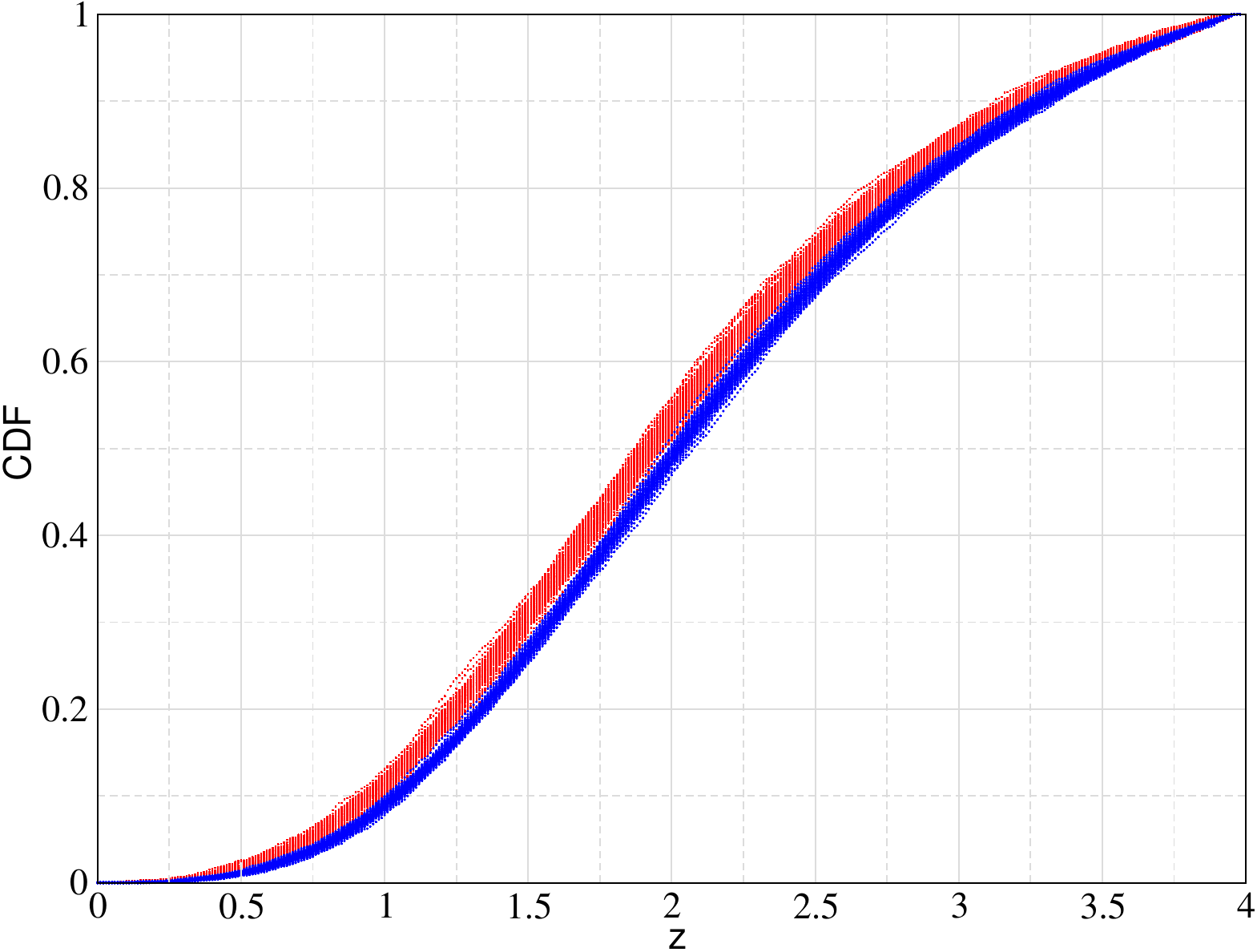}
%\caption{100 GHz}
%\label{fig:depfracz=100}
\end{subfigure}
\caption{CDF of BBH mergers as a function of redshift. We consider two cases as described in the text and sample their redshift distribution using Eq. \ref{eq:BBH_rate_today}. We consider 5000 objects for a given realization and plot several of these realizations. We consider detected BBH mergers until $z=1.4$ (left panel) and $z=4$ (right panel).      }
\label{fig:GW_projection}
\end{figure}

 We do a projection on possibility of distinguishing the two cases that we have discussed above. We sample the redshifts of a given number of objects with a probability distribution which is proportional to the right hand side of Eq. \ref{eq:BBH_rate_today}. We generate many realizations and plot their CDF as a function of redshift in Fig. \ref{fig:GW_projection}. We consider two cases where we have a catalogue of BBH mergers until redshift 1.4 and 4. We find that with about a few thousands of BBH mergers, we may be in a position to distinguish the two cases.

%-----------------------------------------------
%----------------------------------------------

%-----------------------------------------------
%----------------------------------------------

%------------------------------------------------
\section{Conclusions}
\label{sec:conclusions}
%------------------------------------------------
In this work, we showcase the importance of galaxy SMF in population studies of a cosmological distribution of FRBs and BBH mergers. Previous works have neglected this aspect in interpreting their results and have used the observed global SFRD and SMD which is just a function of redshift. In doing so, it is implicitly assumed that all galaxies can potentially act as hosts. However, at least for FRBs, empirical observations point to the possibility of host galaxies with stellar mass peaked around $10^{10}$ $M_{\odot}$. \color{black} With the simplifying assumption that FRBs form in galaxies with a specific stellar mass distribution, \color{black} we find that the relevant SMD with these galaxies is a factor of 3-4 less than the SMD used in previous studies at $z\approx 0$. Therefore, in order to match the observed volumetric rate of FRBs, these host galaxies may have to be more efficient in producing FRBs than previously thought. 

We also show that FRB sources do not fully track the stellar mass distribution of galaxies which is in accordance with previous works. We also find that using the typically assumed SFRD fit in population studies  will bias our interpretation of inferred parameters. This is due to the redshift evolution of galaxy SMF which captures independent information as compared to the redshift dependence in standard SFRD fit. In addition, we study a couple of cases with redshift evolution of stellar mass of FRB hosting galaxies. For mild redshift dependence, we do not see any significant deviation from our fiducial case with no redshift evolution. However, we need a bigger sample of localized FRBs at $z\gtrsim 1$ to see if there is a stronger redshift dependence. Our results are not limited to FRBs and generally applicable to population studies using Gamma Ray Bursts \citep{PGSGVNSMCG2016,WP2010,PD2021}, Gravitational wave events \citep{LIGO2023}. We consider a hypothetical scenario where BBH mergers as detected by LIGO collaboration is hosted in similar galaxies as a sample of localized FRBs. We find that such a scenario can reasonably explain the present data well. 
%This opens up an interesting possibility to infer BBH host galaxy properties using GW data alone. 

\color{black} In this work, we have used a simple two template model with a linear combination of SFRD and SMD. This is motivated from the fact that FRBs form in both star-forming as well as quiescent galaxies where there is not much star formation. In principle, formation of astrophysical objects can have significant correlation with other galaxy properties such as metallicity etc. Empirically, it is seen that metallicity has a simple dependence with stellar mass (figure 10 of \cite{Langeroodi2023} as an example). Therefore, in principle, the metallicity dependence can be captured in our choice of filter function. In order to extract the filter function as well as the weights ($f_Y$) of our templates from data, one needs a larger dataset of well-localized FRBs than presently available. At present, we have only about $\sim 100$ localized FRBs at $z\lesssim 0.5$. As we detect more and more localized FRBs especially at higher redshifts, we may be in a position to test the validity of our two template model. 

\color{black}

Moving forward, in order to use the galaxy SMF with full quantitative confidence, the difference of predicted SMD using this approach and the direct measurements captured in MD fit has to be sorted out. There is a factor of $\approx$ 2 difference between the two values at $z\lesssim 2$ with some redshift dependence. This can potentially bias our parameter inference. With ongoing galaxy surveys such as EUCLID and JWST, we expect to make some progress is resolving these systematic issues.

\section*{Acknowledgements}
We acknowledge discussions with Paz Beniamini throughout this work. SKA was supported by the ARCO fellowship during the initial part of this work.

{%\small
\vspace{-3mm}
\bibliographystyle{unsrtads}
\bibliography{main}

@ARTICLE{BK2025B,
       author = {{Beniamini}, Paz and {Kumar}, Pawan},
        title = "{Can repeating and non-repeating FRBs be drawn from the same population?}",
      journal = {arXiv e-prints},
         year = 2025,
        month = jun,
          eid = {arXiv:2506.09138},
        pages = {arXiv:2506.09138},
          doi = {10.48550/arXiv.2506.09138},
archivePrefix = {arXiv},
       eprint = {2506.09138},
 primaryClass = {astro-ph.HE},
       adsurl = {https://ui.adsabs.harvard.edu/abs/2025arXiv250609138B}
}

@ARTICLE{BK2025,
       author = {{Beniamini}, Paz and {Kumar}, Pawan},
        title = "{The Role of Magnetic and Rotation Axis Alignment in Driving Fast Radio Burst Phenomenology}",
      journal = {\apj},
         year = 2025,
        month = mar,
       volume = {982},
       number = {1},
          eid = {45},
        pages = {45},
          doi = {10.3847/1538-4357/adb8e6},
archivePrefix = {arXiv},
       eprint = {2410.19043},
 primaryClass = {astro-ph.HE},
       adsurl = {https://ui.adsabs.harvard.edu/abs/2025ApJ...982...45B}
}

@ARTICLE{CHIME2020,
	author = {{CHIME/FRB Collaboration} and { Andersen et al }},
	title = "{A bright millisecond-duration radio burst from a Galactic magnetar}",
	journal = {\nat},
	year = 2020,
	month = nov,
	volume = {587},
	number = {7832},
	pages = {54-58},
	doi = {10.1038/s41586-020-2863-y},
	archivePrefix = {arXiv},
	eprint = {2005.10324},
	primaryClass = {astro-ph.HE},
	adsurl = {https://ui.adsabs.harvard.edu/abs/2020Natur.587...54T}
}

@ARTICLE{Totani2023,
       author = {{Totani}, Tomonori and {Tsuzuki}, Yuya},
        title = "{Fast radio bursts trigger aftershocks resembling earthquakes, but not solar flares}",
      journal = {\mnras},
         year = 2023,
        month = dec,
       volume = {526},
       number = {2},
        pages = {2795-2811},
          doi = {10.1093/mnras/stad2532},
archivePrefix = {arXiv},
       eprint = {2306.13612},
 primaryClass = {astro-ph.HE},
       adsurl = {https://ui.adsabs.harvard.edu/abs/2023MNRAS.526.2795T}
}

@ARTICLE{Wadiasingh2019,
       author = {{Wadiasingh}, Zorawar and {Timokhin}, Andrey},
        title = "{Repeating Fast Radio Bursts from Magnetars with Low Magnetospheric Twist}",
      journal = {\apj},
         year = 2019,
        month = jul,
       volume = {879},
       number = {1},
          eid = {4},
        pages = {4},
          doi = {10.3847/1538-4357/ab2240},
archivePrefix = {arXiv},
       eprint = {1904.12036},
 primaryClass = {astro-ph.HE},
       adsurl = {https://ui.adsabs.harvard.edu/abs/2019ApJ...879....4W}
}

@ARTICLE{STARE2020,
	author = {{Bochenek}, C.~D. and {Ravi}, V. and {Belov}, K.~V. and {Hallinan}, G. and
	{Kocz}, J. and {Kulkarni}, S.~R. and {McKenna}, D.~L.},
	title = "{A fast radio burst associated with a Galactic magnetar}",
	journal = {\nat},
	year = 2020,
	month = nov,
	volume = {587},
	number = {7832},
	pages = {59-62},
	doi = {10.1038/s41586-020-2872-x},
	archivePrefix = {arXiv},
	eprint = {2005.10828},
	primaryClass = {astro-ph.HE},
	adsurl = {https://ui.adsabs.harvard.edu/abs/2020Natur.587...59B}
}

@ARTICLE{Bhandari2022,
       author = {{Bhandari et al}, Shivani },
        title = "{Characterizing the Fast Radio Burst Host Galaxy Population and its Connection to Transients in the Local and Extragalactic Universe}",
      journal = {\aj},
         year = 2022,
        month = feb,
       volume = {163},
       number = {2},
          eid = {69},
        pages = {69},
          doi = {10.3847/1538-3881/ac3aec},
archivePrefix = {arXiv},
       eprint = {2108.01282},
 primaryClass = {astro-ph.HE},
       adsurl = {https://ui.adsabs.harvard.edu/abs/2022AJ....163...69B}
}

@ARTICLE{Bannister2019,
       author = {{Bannister et al}, K.~W. },
        title = "{A single fast radio burst localized to a massive galaxy at cosmological distance}",
      journal = {Science},
         year = 2019,
        month = aug,
       volume = {365},
       number = {6453},
        pages = {565-570},
          doi = {10.1126/science.aaw5903},
archivePrefix = {arXiv},
       eprint = {1906.11476},
 primaryClass = {astro-ph.HE},
       adsurl = {https://ui.adsabs.harvard.edu/abs/2019Sci...365..565B}
}

@ARTICLE{Ravi2019,
       author = {{Ravi}, V. and {Catha}, M. and {D'Addario}, L. and {Djorgovski}, S.~G. and {Hallinan}, G. and {Hobbs}, R. and {Kocz}, J. and {Kulkarni}, S.~R. and {Shi}, J. and {Vedantham}, H.~K. and {Weinreb}, S. and {Woody}, D.~P.},
        title = "{A fast radio burst localized to a massive galaxy}",
      journal = {\nat},
         year = 2019,
        month = aug,
       volume = {572},
       number = {7769},
        pages = {352-354},
          doi = {10.1038/s41586-019-1389-7},
archivePrefix = {arXiv},
       eprint = {1907.01542},
 primaryClass = {astro-ph.HE},
       adsurl = {https://ui.adsabs.harvard.edu/abs/2019Natur.572..352R}
}

@ARTICLE{Beniamini2021,
       author = {{Beniamini}, Paz and {Kumar}, Pawan and {Ma}, Xiangcheng and {Quataert}, Eliot},
        title = "{Exploring the epoch of hydrogen reionization using FRBs}",
      journal = {\mnras},
         year = 2021,
        month = apr,
       volume = {502},
       number = {4},
        pages = {5134-5146},
          doi = {10.1093/mnras/stab309},
archivePrefix = {arXiv},
       eprint = {2011.11643},
 primaryClass = {astro-ph.CO},
       adsurl = {https://ui.adsabs.harvard.edu/abs/2021MNRAS.502.5134B}
}

@ARTICLE{James2022,
       author = {{James}, C.~W. and {Ghosh}, E.~M. and {Prochaska}, J.~X. and {Bannister}, K.~W. and {Bhandari}, S. and {Day}, C.~K. and {Deller}, A.~T. and {Glowacki}, M. and {Gordon}, A.~C. and {Heintz}, K.~E. and {Marnoch}, L. and {Ryder}, S.~D. and {Scott}, D.~R. and {Shannon}, R.~M. and {Tejos}, N.},
        title = "{A measurement of Hubble's Constant using Fast Radio Bursts}",
      journal = {\mnras},
         year = 2022,
        month = nov,
       volume = {516},
       number = {4},
        pages = {4862-4881},
          doi = {10.1093/mnras/stac2524},
archivePrefix = {arXiv},
       eprint = {2208.00819},
 primaryClass = {astro-ph.CO},
       adsurl = {https://ui.adsabs.harvard.edu/abs/2022MNRAS.516.4862J}
}

@ARTICLE{Sharma2024,
       author = {{Sharma et al}, Kritti },
        title = "{Preferential Occurrence of Fast Radio Bursts in Massive Star-Forming Galaxies}",
      journal = {arXiv e-prints},
         year = 2024,
        month = sep,
          eid = {arXiv:2409.16964},
        pages = {arXiv:2409.16964},
          doi = {10.48550/arXiv.2409.16964},
archivePrefix = {arXiv},
       eprint = {2409.16964},
 primaryClass = {astro-ph.HE},
       adsurl = {https://ui.adsabs.harvard.edu/abs/2024arXiv240916964S}
}

@ARTICLE{Connor2024,
       author = {{Connor}, Liam and {Ravi}, Vikram and {Sharma}, Kritti and {Ocker}, Stella Koch and {Faber}, Jakob and {Hallinan}, Gregg and {Harnach}, Charlie and {Hellbourg}, Greg and {Hobbs}, Rick and {Hodge}, David and {Hodges}, Mark and {Kosogorov}, Nikita and {Lamb}, James and {Law}, Casey and {Rasmussen}, Paul and {Sherman}, Myles and {Somalwar}, Jean and {Weinreb}, Sander and {Woody}, David},
        title = "{A gas rich cosmic web revealed by partitioning the missing baryons}",
      journal = {arXiv e-prints},
         year = 2024,
        month = sep,
          eid = {arXiv:2409.16952},
        pages = {arXiv:2409.16952},
          doi = {10.48550/arXiv.2409.16952},
archivePrefix = {arXiv},
       eprint = {2409.16952},
 primaryClass = {astro-ph.CO},
       adsurl = {https://ui.adsabs.harvard.edu/abs/2024arXiv240916952C}
}

@ARTICLE{Gordon2023_1,
       author = {{Gordon et al}, Alexa C. },
        title = "{The Demographics, Stellar Populations, and Star Formation Histories of Fast Radio Burst Host Galaxies: Implications for the Progenitors}",
      journal = {\apj},
         year = 2023,
        month = sep,
       volume = {954},
       number = {1},
          eid = {80},
        pages = {80},
          doi = {10.3847/1538-4357/ace5aa},
archivePrefix = {arXiv},
       eprint = {2302.05465},
 primaryClass = {astro-ph.GA},
       adsurl = {https://ui.adsabs.harvard.edu/abs/2023ApJ...954...80G}
}

@ARTICLE{AB2025,
       author = {{Acharya}, Sandeep Kumar and {Beniamini}, Paz},
        title = "{Redshift dependence of FRB host dispersion measures across cosmic epochs}",
      journal = {\jcap},
         year = 2025,
        month = jan,
       volume = {2025},
       number = {1},
          eid = {036},
        pages = {036},
          doi = {10.1088/1475-7516/2025/01/036},
archivePrefix = {arXiv},
       eprint = {2408.03163},
 primaryClass = {astro-ph.CO},
       adsurl = {https://ui.adsabs.harvard.edu/abs/2025JCAP...01..036A}
}

@ARTICLE{Chime_catalog,
       author = {{CHIME/FRB Collaboration} and {Amiri et al} },
        title = "{The First CHIME/FRB Fast Radio Burst Catalog}",
      journal = {\apjs},
         year = 2021,
        month = dec,
       volume = {257},
       number = {2},
          eid = {59},
        pages = {59},
          doi = {10.3847/1538-4365/ac33ab},
archivePrefix = {arXiv},
       eprint = {2106.04352},
 primaryClass = {astro-ph.HE},
       adsurl = {https://ui.adsabs.harvard.edu/abs/2021ApJS..257...59C}
}

@ARTICLE{BK2023,
       author = {{Beniamini}, Paz and {Kumar}, Pawan},
        title = "{Hybrid pulsar-magnetar model for FRB 20191221A}",
      journal = {\mnras},
         year = 2023,
        month = mar,
       volume = {519},
       number = {4},
        pages = {5345-5351},
          doi = {10.1093/mnras/stad028},
archivePrefix = {arXiv},
       eprint = {2211.07669},
 primaryClass = {astro-ph.HE},
       adsurl = {https://ui.adsabs.harvard.edu/abs/2023MNRAS.519.5345B}
}

@ARTICLE{2025arXiv250215566L,
       author = {{Loudas}, Nick and {Li}, Dongzi and {Strauss}, Michael A. and {Leja}, Joel},
        title = "{Unveiling the origin of fast radio bursts by modeling the stellar mass and star formation distributions of their host galaxies}",
      journal = {arXiv e-prints},
         year = 2025,
        month = feb,
          eid = {arXiv:2502.15566},
        pages = {arXiv:2502.15566},
          doi = {10.48550/arXiv.2502.15566},
archivePrefix = {arXiv},
       eprint = {2502.15566},
 primaryClass = {astro-ph.HE},
       adsurl = {https://ui.adsabs.harvard.edu/abs/2025arXiv250215566L}
}

@ARTICLE{Cheng+20,
       author = {{Cheng}, Yingjie and {Zhang}, G.~Q. and {Wang}, F.~Y.},
        title = "{Statistical properties of magnetar bursts and FRB 121102}",
      journal = {\mnras},
         year = 2020,
        month = jan,
       volume = {491},
       number = {1},
        pages = {1498-1505},
          doi = {10.1093/mnras/stz3085},
archivePrefix = {arXiv},
       eprint = {1910.14201},
 primaryClass = {astro-ph.HE},
       adsurl = {https://ui.adsabs.harvard.edu/abs/2020MNRAS.491.1498C}
}

@ARTICLE{Shah2025,
       author = {{Shah et al}, Vishwangi },
        title = "{A Repeating Fast Radio Burst Source in the Outskirts of a Quiescent Galaxy}",
      journal = {\apjl},
         year = 2025,
        month = feb,
       volume = {979},
       number = {2},
          eid = {L21},
        pages = {L21},
          doi = {10.3847/2041-8213/ad9ddc},
archivePrefix = {arXiv},
       eprint = {2410.23374},
 primaryClass = {astro-ph.HE},
       adsurl = {https://ui.adsabs.harvard.edu/abs/2025ApJ...979L..21S}
}

@ARTICLE{Eftekhari2025,
       author = {{Eftekhari et al}, T. },
        title = "{The Massive and Quiescent Elliptical Host Galaxy of the Repeating Fast Radio Burst FRB 20240209A}",
      journal = {\apjl},
         year = 2025,
        month = feb,
       volume = {979},
       number = {2},
          eid = {L22},
        pages = {L22},
          doi = {10.3847/2041-8213/ad9de2},
archivePrefix = {arXiv},
       eprint = {2410.23336},
 primaryClass = {astro-ph.HE},
       adsurl = {https://ui.adsabs.harvard.edu/abs/2025ApJ...979L..22E}
}

@ARTICLE{Kirsten2022,
       author = {{Kirsten et al}, F. },
        title = "{A repeating fast radio burst source in a globular cluster}",
      journal = {\nat},
         year = 2022,
        month = feb,
       volume = {602},
       number = {7898},
        pages = {585-589},
          doi = {10.1038/s41586-021-04354-w},
archivePrefix = {arXiv},
       eprint = {2105.11445},
 primaryClass = {astro-ph.HE},
       adsurl = {https://ui.adsabs.harvard.edu/abs/2022Natur.602..585K}
}

@ARTICLE{Gupta2025,
       author = {{Gupta}, Om and {Beniamini}, Paz and {Kumar}, Pawan and {Finkelstein}, Steven L.},
        title = "{The cosmic evolution of FRBs inferred from CHIME/FRB Catalog 1}",
      journal = {arXiv e-prints},
         year = 2025,
        month = jan,
          eid = {arXiv:2501.09810},
        pages = {arXiv:2501.09810},
          doi = {10.48550/arXiv.2501.09810},
archivePrefix = {arXiv},
       eprint = {2501.09810},
 primaryClass = {astro-ph.HE},
       adsurl = {https://ui.adsabs.harvard.edu/abs/2025arXiv250109810G}
}

@ARTICLE{Shannon2024,
       author = {{Shannon et al}, R.~M. },
        title = "{The Commensal Real-time ASKAP Fast Transient incoherent-sum survey}",
      journal = {arXiv e-prints},
         year = 2024,
        month = aug,
          eid = {arXiv:2408.02083},
        pages = {arXiv:2408.02083},
          doi = {10.48550/arXiv.2408.02083},
archivePrefix = {arXiv},
       eprint = {2408.02083},
 primaryClass = {astro-ph.HE},
       adsurl = {https://ui.adsabs.harvard.edu/abs/2024arXiv240802083S}
}

@ARTICLE{Euclid2025,
       author = {{Euclid Collaboration} },
        title = "{Euclid preparation. Cosmic Dawn Survey: evolution of the galaxy stellar mass function across $0.2<z<6.5$ measured over 10 square degrees}",
      journal = {arXiv e-prints},
         year = 2025,
        month = apr,
          eid = {arXiv:2504.17867},
        pages = {arXiv:2504.17867},
          doi = {10.48550/arXiv.2504.17867},
archivePrefix = {arXiv},
       eprint = {2504.17867},
 primaryClass = {astro-ph.GA},
       adsurl = {https://ui.adsabs.harvard.edu/abs/2025arXiv250417867E}
}

@ARTICLE{MD2014,
       author = {{Madau}, Piero and {Dickinson}, Mark},
        title = "{Cosmic Star-Formation History}",
      journal = {\araa},
         year = 2014,
        month = aug,
       volume = {52},
        pages = {415-486},
          doi = {10.1146/annurev-astro-081811-125615},
archivePrefix = {arXiv},
       eprint = {1403.0007},
 primaryClass = {astro-ph.CO},
       adsurl = {https://ui.adsabs.harvard.edu/abs/2014ARA&A..52..415M}
}

@ARTICLE{HM2025,
       author = {{Horowicz}, Asaf and {Margalit}, Ben},
        title = "{The Host Galaxies of Fast Radio Bursts Track a Combination of Stellar Mass and Star Formation, Similar to Type Ia Supernovae}",
      journal = {arXiv e-prints},
         year = 2025,
        month = apr,
          eid = {arXiv:2504.08038},
        pages = {arXiv:2504.08038},
          doi = {10.48550/arXiv.2504.08038},
archivePrefix = {arXiv},
       eprint = {2504.08038},
 primaryClass = {astro-ph.HE},
       adsurl = {https://ui.adsabs.harvard.edu/abs/2025arXiv250408038H}
}

@ARTICLE{COSMOSWEB2025,
       author = {{Shuntov et al}, M. },
        title = "{COSMOS-Web: Stellar mass assembly in relation to dark matter halos across 0.2 < z < 12 of cosmic history}",
      journal = {\aap},
         year = 2025,
        month = mar,
       volume = {695},
          eid = {A20},
        pages = {A20},
          doi = {10.1051/0004-6361/202452570},
archivePrefix = {arXiv},
       eprint = {2410.08290},
 primaryClass = {astro-ph.GA},
       adsurl = {https://ui.adsabs.harvard.edu/abs/2025A&A...695A..20S}
}

@ARTICLE{HB2006,
       author = {{Hopkins}, Andrew M. and {Beacom}, John F.},
        title = "{On the Normalization of the Cosmic Star Formation History}",
      journal = {\apj},
         year = 2006,
        month = nov,
       volume = {651},
       number = {1},
        pages = {142-154},
          doi = {10.1086/506610},
archivePrefix = {arXiv},
       eprint = {astro-ph/0601463},
 primaryClass = {astro-ph},
       adsurl = {https://ui.adsabs.harvard.edu/abs/2006ApJ...651..142H}
}

@ARTICLE{WTH2008,
       author = {{Wilkins}, Stephen M. and {Trentham}, Neil and {Hopkins}, Andrew M.},
        title = "{The evolution of stellar mass and the implied star formation history}",
      journal = {\mnras},
         year = 2008,
        month = apr,
       volume = {385},
       number = {2},
        pages = {687-694},
          doi = {10.1111/j.1365-2966.2008.12885.x},
archivePrefix = {arXiv},
       eprint = {0801.1594},
 primaryClass = {astro-ph},
       adsurl = {https://ui.adsabs.harvard.edu/abs/2008MNRAS.385..687W}
}

@INPROCEEDINGS{PP2010,
       author = {{Popov}, Sergey B. and {Postnov}, K.~A.},
        title = "{Hyperflares of SGRs as an engine for millisecond extragalactic radio bursts}",
    booktitle = {Evolution of Cosmic Objects through their Physical Activity},
         year = 2010,
       editor = {{Harutyunian}, H.~A. and {Mickaelian}, A.~M. and {Terzian}, Y.},
        month = nov,
        pages = {129-132},
          doi = {10.48550/arXiv.0710.2006},
archivePrefix = {arXiv},
       eprint = {0710.2006},
 primaryClass = {astro-ph},
       adsurl = {https://ui.adsabs.harvard.edu/abs/2010vaoa.conf..129P}
}

@ARTICLE{KLB2017,
       author = {{Kumar}, Pawan and {Lu}, Wenbin and {Bhattacharya}, Mukul},
        title = "{Fast radio burst source properties and curvature radiation model}",
      journal = {\mnras},
         year = 2017,
        month = jul,
       volume = {468},
       number = {3},
        pages = {2726-2739},
          doi = {10.1093/mnras/stx665},
archivePrefix = {arXiv},
       eprint = {1703.06139},
 primaryClass = {astro-ph.HE},
       adsurl = {https://ui.adsabs.harvard.edu/abs/2017MNRAS.468.2726K}
}

@ARTICLE{Bhardwaj2021_1,
       author = {{Bhardwaj}, M. and {Gaensler}, B.~M. and {Kaspi}, V.~M. and {Landecker}, T.~L. and {Mckinven}, R. and {Michilli}, D. and {Pleunis}, Z. and {Tendulkar}, S.~P. and {Andersen}, B.~C. and {Boyle}, P.~J. and {Cassanelli}, T. and {Chawla}, P. and {Cook}, A. and {Dobbs}, M. and {Fonseca}, E. and {Kaczmarek}, J. and {Leung}, C. and {Masui}, K. and {Mnchmeyer}, M. and {Ng}, C. and {Rafiei-Ravandi}, M. and {Scholz}, P. and {Shin}, K. and {Smith}, K.~M. and {Stairs}, I.~H. and {Zwaniga}, A.~V.},
        title = "{A Nearby Repeating Fast Radio Burst in the Direction of M81}",
      journal = {\apjl},
         year = 2021,
        month = apr,
       volume = {910},
       number = {2},
          eid = {L18},
        pages = {L18},
          doi = {10.3847/2041-8213/abeaa6},
archivePrefix = {arXiv},
       eprint = {2103.01295},
 primaryClass = {astro-ph.HE},
       adsurl = {https://ui.adsabs.harvard.edu/abs/2021ApJ...910L..18B}
}

@ARTICLE{ZZLL2021,
       author = {{Zhang}, Rachel C. and {Zhang}, Bing and {Li}, Ye and {Lorimer}, Duncan R.},
        title = "{On the energy and redshift distributions of fast radio bursts}",
      journal = {\mnras},
         year = 2021,
        month = feb,
       volume = {501},
       number = {1},
        pages = {157-167},
          doi = {10.1093/mnras/staa3537},
archivePrefix = {arXiv},
       eprint = {2011.06151},
 primaryClass = {astro-ph.HE},
       adsurl = {https://ui.adsabs.harvard.edu/abs/2021MNRAS.501..157Z}
}

@ARTICLE{James2022_1,
       author = {{James}, C.~W. and {Prochaska}, J.~X. and {Macquart}, J. -P. and {North-Hickey}, F.~O. and {Bannister}, K.~W. and {Dunning}, A.},
        title = "{The fast radio burst population evolves, consistent with the star formation rate}",
      journal = {\mnras},
         year = 2022,
        month = feb,
       volume = {510},
       number = {1},
        pages = {L18-L23},
          doi = {10.1093/mnrasl/slab117},
archivePrefix = {arXiv},
       eprint = {2101.07998},
 primaryClass = {astro-ph.HE},
       adsurl = {https://ui.adsabs.harvard.edu/abs/2022MNRAS.510L..18J}
}

@ARTICLE{Chen2025,
       author = {{Chen}, Xiang-Lei and {Tsai}, Chao-Wei and {Stern}, Daniel and {Bochenek}, Christopher D. and {Chatterjee}, Shami and {Law}, Casey and {Li}, Di and {Niu}, Chen-hui and {Niino}, Yuu and {Feng}, Yi and {Wang}, Pei and {Assef}, Roberto J. and {Li}, Guo-dong and {Lake}, Sean E. and {Luo}, Gan and {Liao}, Mai},
        title = "{The Host Galaxy of FRB 20190520B and Its Unique Ionized Gas Distribution}",
      journal = {\apj},
         year = 2025,
        month = apr,
       volume = {982},
       number = {2},
          eid = {203},
        pages = {203},
          doi = {10.3847/1538-4357/adb84d},
archivePrefix = {arXiv},
       eprint = {2503.01740},
 primaryClass = {astro-ph.GA},
       adsurl = {https://ui.adsabs.harvard.edu/abs/2025ApJ...982..203C}
}

@ARTICLE{Gordon2025,
       author = {{Gordon}, Alexa C. and {Fong}, Wen-fai and {Deller}, Adam T. and {Marnoch}, Lachlan and {Lim}, Sungsoon and {Peng}, Eric W. and {Bannister}, Keith W. and {Bera}, Apurba and {Bhat}, N.~D.~R. and {Dial}, Tyson and {Dong}, Yuxin and {Eftekhari}, Tarraneh and {Glowacki}, Marcin and {Gourdji}, Kelly and {Gupta}, Vivek and {Jahns-Schindler}, Joscha N. and {Jaini}, Akhil and {Kilpatrick}, Charles D. and {Liu}, Chang and {Prochaska}, J. Xavier and {Ryder}, Stuart D. and {Shannon}, Ryan M. and {Simha}, Sunil and {Tejos}, Nicolas and {Wang}, Yuanming and {Wang}, Ziteng},
        title = "{Mapping the Spatial Distribution of Fast Radio Bursts within their Host Galaxies}",
      journal = {arXiv e-prints},
         year = 2025,
        month = jun,
          eid = {arXiv:2506.06453},
        pages = {arXiv:2506.06453},
          doi = {10.48550/arXiv.2506.06453},
archivePrefix = {arXiv},
       eprint = {2506.06453},
 primaryClass = {astro-ph.GA},
       adsurl = {https://ui.adsabs.harvard.edu/abs/2025arXiv250606453G}
}

@ARTICLE{PGSGVNSMCG2016,
       author = {{Pescalli}, A. and {Ghirlanda}, G. and {Salvaterra}, R. and {Ghisellini}, G. and {Vergani}, S.~D. and {Nappo}, F. and {Salafia}, O.~S. and {Melandri}, A. and {Covino}, S. and {G{\"o}tz}, D.},
        title = "{The rate and luminosity function of long gamma ray bursts}",
      journal = {\aap},
         year = 2016,
        month = mar,
       volume = {587},
          eid = {A40},
        pages = {A40},
          doi = {10.1051/0004-6361/201526760},
archivePrefix = {arXiv},
       eprint = {1506.05463},
 primaryClass = {astro-ph.HE},
       adsurl = {https://ui.adsabs.harvard.edu/abs/2016A&A...587A..40P}
}

@ARTICLE{WP2010,
       author = {{Wanderman}, David and {Piran}, Tsvi},
        title = "{The luminosity function and the rate of Swift's gamma-ray bursts}",
      journal = {\mnras},
         year = 2010,
        month = aug,
       volume = {406},
       number = {3},
        pages = {1944-1958},
          doi = {10.1111/j.1365-2966.2010.16787.x},
archivePrefix = {arXiv},
       eprint = {0912.0709},
 primaryClass = {astro-ph.HE},
       adsurl = {https://ui.adsabs.harvard.edu/abs/2010MNRAS.406.1944W}
}

@ARTICLE{PD2021,
       author = {{Palmerio}, J.~T. and {Daigne}, F.},
        title = "{Constraining the intrinsic population of long gamma-ray bursts: Implications for spectral correlations, cosmic evolution, and their use as tracers of star formation}",
      journal = {\aap},
         year = 2021,
        month = may,
       volume = {649},
          eid = {A166},
        pages = {A166},
          doi = {10.1051/0004-6361/202039929},
archivePrefix = {arXiv},
       eprint = {2011.14745},
 primaryClass = {astro-ph.HE},
       adsurl = {https://ui.adsabs.harvard.edu/abs/2021A&A...649A.166P}
}

@ARTICLE{Shin2023,
       author = {{Shin}, Kaitlyn and {Masui}, Kiyoshi W. and {Bhardwaj}, Mohit and {Cassanelli}, Tomas and {Chawla}, Pragya and {Dobbs}, Matt and {Dong}, Fengqiu Adam and {Fonseca}, Emmanuel and {Gaensler}, B.~M. and {Herrera-Mart{\'\i}n}, Antonio and {Kaczmarek}, Jane and {Kaspi}, Victoria and {Leung}, Calvin and {Merryfield}, Marcus and {Michilli}, Daniele and {M{\"u}nchmeyer}, Moritz and {Pearlman}, Aaron B. and {Rafiei-Ravandi}, Masoud and {Smith}, Kendrick and {Stairs}, Ingrid and {Tendulkar}, Shriharsh P.},
        title = "{Inferring the Energy and Distance Distributions of Fast Radio Bursts Using the First CHIME/FRB Catalog}",
      journal = {\apj},
         year = 2023,
        month = feb,
       volume = {944},
       number = {1},
          eid = {105},
        pages = {105},
          doi = {10.3847/1538-4357/acaf06},
archivePrefix = {arXiv},
       eprint = {2207.14316},
 primaryClass = {astro-ph.HE},
       adsurl = {https://ui.adsabs.harvard.edu/abs/2023ApJ...944..105S}
}

@ARTICLE{Vijaykumar2024,
       author = {{Vijaykumar}, Aditya and {Fishbach}, Maya and {Adhikari}, Susmita and {Holz}, Daniel E.},
        title = "{Inferring Host-galaxy Properties of LIGO{\textendash}Virgo{\textendash}KAGRA's Black Holes}",
      journal = {\apj},
         year = 2024,
        month = sep,
       volume = {972},
       number = {2},
          eid = {157},
        pages = {157},
          doi = {10.3847/1538-4357/ad6140},
archivePrefix = {arXiv},
       eprint = {2312.03316},
 primaryClass = {astro-ph.HE},
       adsurl = {https://ui.adsabs.harvard.edu/abs/2024ApJ...972..157V}
}

@ARTICLE{LIGO2023,
       author = {{Abbott et. al.}, R.},
        title = "{Population of Merging Compact Binaries Inferred Using Gravitational Waves through GWTC-3}",
      journal = {Physical Review X},
         year = 2023,
        month = jan,
       volume = {13},
       number = {1},
          eid = {011048},
        pages = {011048},
          doi = {10.1103/PhysRevX.13.011048},
archivePrefix = {arXiv},
       eprint = {2111.03634},
 primaryClass = {astro-ph.HE},
       adsurl = {https://ui.adsabs.harvard.edu/abs/2023PhRvX..13a1048A}
}

@ARTICLE{Langeroodi2023,
       author = {{Langeroodi}, Danial and {Hjorth}, Jens and {Chen}, Wenlei and {Kelly}, Patrick L. and {Williams}, Hayley and {Lin}, Yu-Heng and {Scarlata}, Claudia and {Zitrin}, Adi and {Broadhurst}, Tom and {Diego}, Jose M. and {Huang}, Xiaosheng and {Filippenko}, Alexei V. and {Foley}, Ryan J. and {Jha}, Saurabh and {Koekemoer}, Anton M. and {Oguri}, Masamune and {Perez-Fournon}, Ismael and {Pierel}, Justin and {Poidevin}, Frederick and {Strolger}, Lou},
        title = "{Evolution of the Mass-Metallicity Relation from Redshift z {\ensuremath{\approx}} 8 to the Local Universe}",
      journal = {\apj},
         year = 2023,
        month = nov,
       volume = {957},
       number = {1},
          eid = {39},
        pages = {39},
          doi = {10.3847/1538-4357/acdbc1},
archivePrefix = {arXiv},
       eprint = {2212.02491},
 primaryClass = {astro-ph.GA},
       adsurl = {https://ui.adsabs.harvard.edu/abs/2023ApJ...957...39L}
}
}

\appendix

%--------------------------------------------------------
\section{Parameters used and least square fitting for Stellar Mass function (SMF)}
\label{app:SMF_fit}
%--------------------------------------------------------
In table \ref{tab:parameter_fit}, we tabulate the best fit parameters of Schechter functions obtained in \cite{Euclid2025}. We choose the midpoint of the redshift bin as our value. The authors in \cite{Euclid2025} fit a double and single Schechter function to the galaxy SMF for $z\lesssim 2$ and $z\gtrsim 2$ respectively. We do a least square fitting using the tabulated values to obtain smooth, continuous results over $z$. We do least square parabola fitting for ${\rm log10(\mathcal M_*)}, \alpha_1$ and ${\rm log10}(\Phi_1)$ and line fitting for $\alpha_2$ and ${\rm log10}(\Phi_2)$. We use the double Schechter function with extrapolation of our least square fit. We find that this procedure provides the best fit to the data as shown in Fig. \ref{fig:euclid_data_fit}. A single Schechter function provides a worse fit to data at $z\lesssim 2$, while a transition from single to double Schechter function creates a sharp jump at $z\approx 2$.

%------------------------------------------------
\begin{table}[H]
  \begin{center}
   \begin{tabular}{l|c|c|c|c|r} % <-- Alignments: 1st column left, 2nd middle and 3rd right, with vertical lines in between
      %\textbf{V} & \textbf{Value 2} & \textbf{Value 3}\\
   z & log10({\it $\mathcal {M*}$}) $M_{\odot}$ & $\alpha_1$ & ${\rm log10}{(\Phi_1)}$ ${\rm Mpc^{-3}dex^{-1}}$ & $\alpha_2$ &  ${\rm log10}{(\Phi_2)}$ ${\rm Mpc^{-3}dex^{-1}}$  \\
    \hline
    0.35 & 10.94 & -1.55 & -3.38 & -0.84 & -2.79 \\
    0.65 & 10.87 & -1.40 & -3.09 & -0.56 & -2.74 \\
    0.95 & 10.89 & -1.54 & -3.46 & -0.55 & -2.93 \\
    1.3 & 10.99 & -1.55 & -3.84 & -0.86 & -3.12 \\
    1.75 & 11.14 & -1.55 & -4.05 & -1.10 & -3.56 \\
    2.25 & 11.15 & -1.55 & -3.84 & - & -  \\
    2.75 & 11.05 & -1.70 & -3.94 & - & -  \\
    3.25 & 11.05 & -1.70 & -4.09 & - & -  \\
    4.0 & 10.93 & -2.05 & -4.57 & - & -  \\
    5.0 & 10.95 & -2.20 & -5.46 & - & -  \\
    6.0 & 11.20 & -2.20 & -6.09 & - & -  \\
 \end{tabular}
 \caption{The parameters used in this work for fitting SMF across redshift. These are derived from Table A.1 of \citep{Euclid2025}. We have used the central value of their redshift bin and the maximum likelihood values of other parameters. We use this table to obtain a least square fit across the redshift range in a continuous and smooth fashion. }
 \label{tab:parameter_fit}
   \end{center}
   \end{table}
%-------------------------------------------------

%--------------------------------------------------
\section{Sample of localized FRBs with stellar mass}
%--------------------------------------------------

%\begin{table}[H]
%  \begin{center}
  \begin{table}[H]
  \begin{minipage}[t]{.5\linewidth}\centering
\begin{tabular}{|c|c|c|}
%   \begin{tabular}{l|c|r} % <-- Alignments: 1st column left, 2nd middle and 3rd right, with vertical lines in between
      %\textbf{V} & \textbf{Value 2} & \textbf{Value 3}\\
   FRB & $z$ & log10($M_*$) $M_{ \odot}$   \\
    \hline
20220319D & 0.0112 & 10.1 \\ 
20180916B & 0.0330 & $9.91_{-0.05}^{+0.03}$ \\
20231120A & 0.0368 & 10.4 \\
20220207C & 0.0433 & $9.95_{-0.03}^{+0.03}$ \\
20211127I & 0.0469 & $9.48_{-0.02}^{+0.06}$ \\
20211212A & 0.0707 & $10.28_{-0.06}^{+0.05}$ \\
20220509G & 0.0894 & $10.7_{-0.01}^{+0.01}$ \\
20230124  & 0.0939 & 9.46 \\
20201124A & 0.0980 & $10.22_{-0.05}^{+0.05}$ \\
20220914A & 0.1139 & $9.24_{-0.04}^{+0.08}$ \\
20190608B & 0.1178 & $10.56_{-0.02}^{+0.02}$ \\
20230628A & 0.1270 & $9.29_{-0.03}^{+0.03}$ \\
 20210804D & 0.1293 & $10.97_{-0.02}^{+0.02}$ \\
20210410D & 0.1415 & $9.47_{-0.05}^{+0.05}$ \\
20220920A & 0.1582 & $9.87_{-0.01}^{+0.01}$ \\
20200430A & 0.1607 & $9.3_{-0.1}^{+0.07}$ \\
20121102A & 0.1931 & $8.14_{-0.1}^{+0.09}$ \\
20210117A & 0.2145 & $8.59_{-0.06}^{+0.05}$ \\
20191001A & 0.2342 & $10.73_{-0.08}^{+0.07}$ \\
20190714A & 0.2365 & $10.22_{-0.04}^{+0.04}$ \\
20221101B & 0.2395 & $11.21_{-0.02}^{+0.03}$ \\
20220825A & 0.2414 & $10.01_{-0.06}^{+0.06}$ \\
20190520B & 0.2417 & $9.08_{-0.09}^{+0.08}$ \\
 20220307B & 0.2481 & $10.14_{-0.04}^{+0.03}$ \\
20221113A & 0.2505 & $9.48_{-0.04}^{+0.04}$ \\
20231123B & 0.2621 & $11.04_{-0.01}^{+0.01}$ \\
20230307A & 0.2706 & $10.76_{-0.02}^{+0.03}$ \\
 \end{tabular}
 \end{minipage}%
 \hfill
% \caption{ }
% \label{}
%   \end{center}
%   \end{table}
%----------------------------------------------
%\begin{table}[hbtp]
%\begin{twocolumn}
%  \begin{center}
%    \begin{table}[hbtp]
 \begin{minipage}[t]{.5\linewidth}
\begin{tabular}{|c|c|c|}
 %  \begin{tabular}{l|c|r} % <-- Alignments: 1st column left, 2nd middle and 3rd right, with vertical lines in between
      %\textbf{V} & \textbf{Value 2} & \textbf{Value 3}\\
  FRB & $z$ & log10($M_*$) $M_{ \odot}$   \\
    \hline  
20221116A & 0.2764 & $11.01_{-0.02}^{0.02}$ \\      
20220105A & 0.2784 & $10.01_{-0.1}^{+0.05}$ \\    
20210320C & 0.2796 & $10.37_{-0.06}^{+0.05}$ \\  
20221012A & 0.2847 & $10.96_{-0.02}^{+0.02}$ \\
20190102C & 0.2909 & $9.69_{-0.11}^{+0.09}$ \\
20220506D & 0.3 & $10.45_{-0.03}^{+0.03}$ \\
20230501A & 0.3015 & 10.29 \\
20180924B & 0.3212 & $10.39_{-0.02}^{+0.02}$ \\
20230626A & 0.3270 & $10.44_{-0.04}^{+0.04}$ \\
20180301A & 0.3305 & $9.64_{-0.11}^{+0.11}$ \\   
20211203C & 0.3437 & $9.76_{-0.09}^{+0.07}$ \\
20220208A & 0.3510 & $10.08_{-0.02}^{+0.02}$ \\
20220726A & 0.3619 & $10.18_{-0.03}^{+0.04}$ \\
20200906A & 0.3688 & $10.37_{-0.05}^{+0.05}$ \\
20220330D & 0.3714 & $10.5_{-0.01}^{+0.02}$ \\
20190611B & 0.3778 & $9.57_{-0.12}^{+0.12}$ \\
20220204A & 0.4012 & $9.7_{-0.09}^{+0.04}$ \\
20230712A & 0.4525 & $11.13_{-0.01}^{+0.01}$ \\
20181112A & 0.4755 & $9.87_{-0.07}^{+0.07}$ \\
20220310F & 0.4780 & $9.98_{-0.06}^{+0.08}$ \\
20190711A & 0.5218 & $9.10_{-0.23}^{+0.15}$ \\
20230216A & 0.5310 & 9.82 \\
20221027A & 0.5422 & $9.47_{-0.04}^{+0.07}$ \\
20221219A & 0.5530 & $10.21_{-0.04}^{+0.03}$ \\
20220418A & 0.6214 & $10.26_{-0.02}^{+0.02}$ \\
20221029A & 0.9750 & $10.59_{-0.10}^{+0.14}$ \\

 \end{tabular}
 \end{minipage}
 \caption{Sample of localized FRBs with their measured stellar mass. This table is compiled from the data in \citep{Gordon2023_1} and \citep{Sharma2024}.  }
 \label{tab:FRB_sample}
%   \end{center}
   \end{table}
%----------------------------------------------    

%----------------------------------------------
\section{Stellar mass distribution of host galaxies for a sample of localized FRB host galaxies}
\label{app:stellarmass_data}
\begin{figure}[!htp]
\begin{subfigure}[b]{0.4\textwidth}
\includegraphics[scale=0.3]{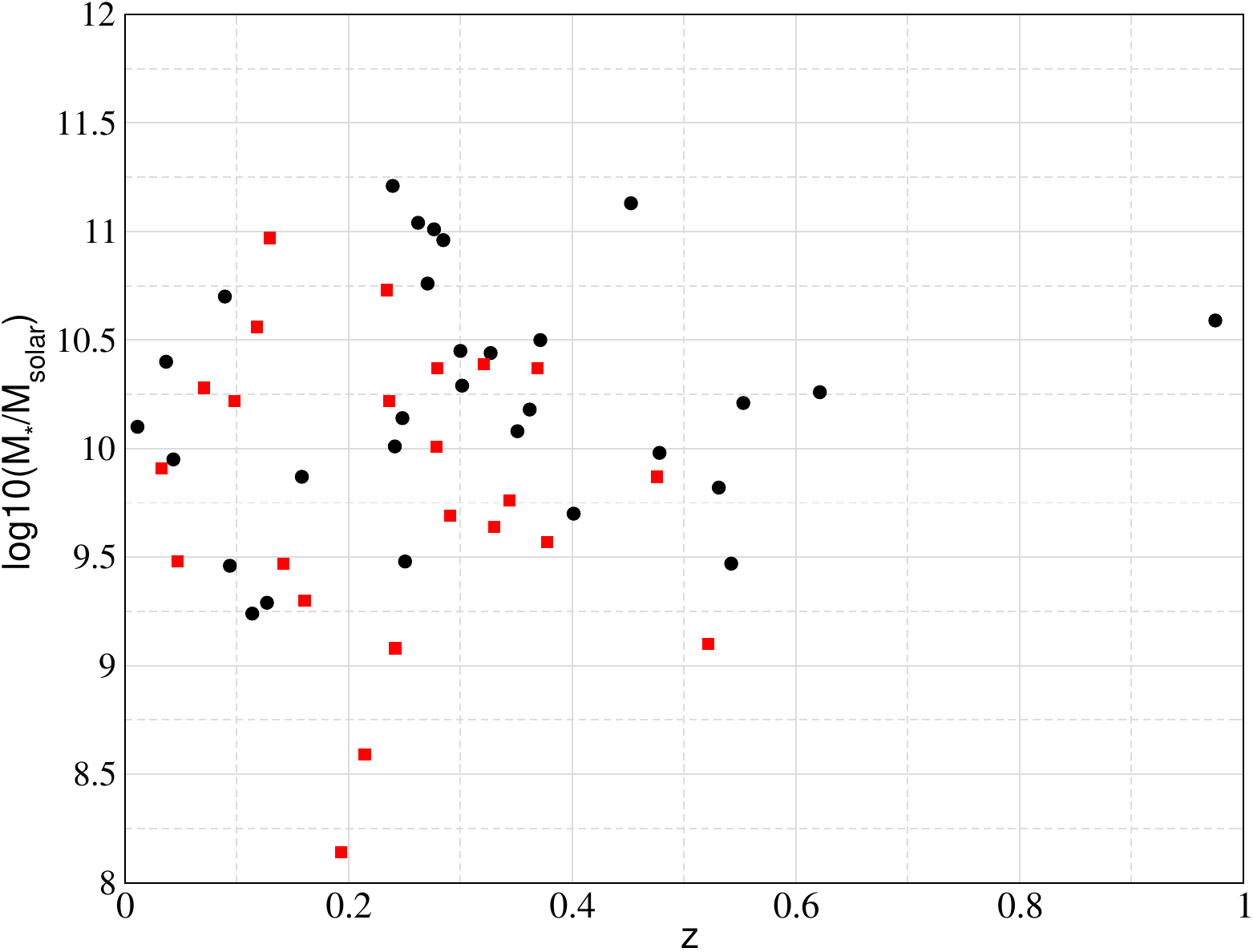}
%\caption{70 GHz}
%\label{fig:depfracz=1000}
\end{subfigure}\hspace{50 pt}
\begin{subfigure}[b]{0.4\textwidth}
\includegraphics[scale=0.3]{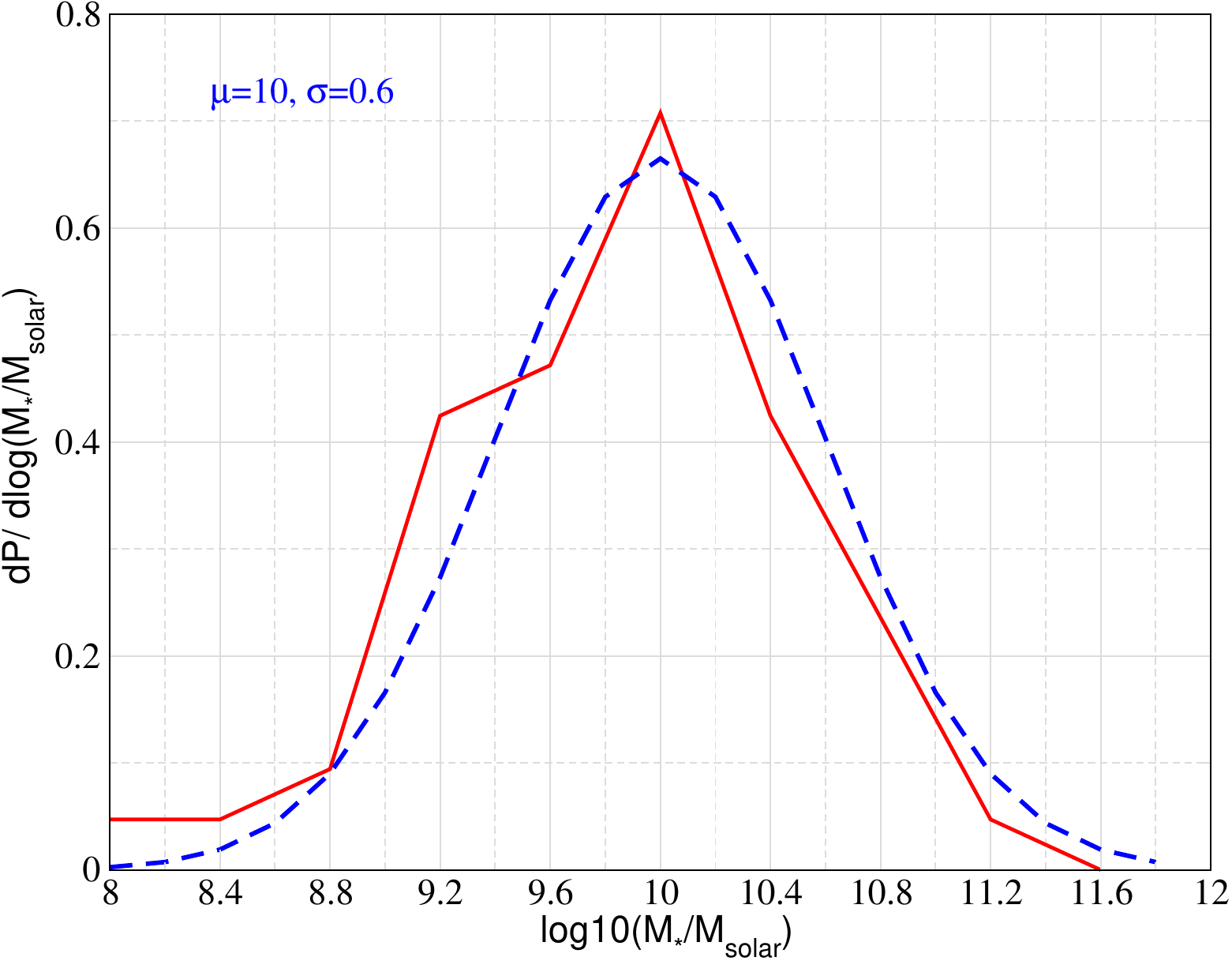}
%\caption{100 GHz}
%\label{fig:depfracz=100}
\end{subfigure}
\caption{Distribution of sample of 53 FRBs used in Table \ref{tab:FRB_sample}. In the left panel, we plot the stellar mass of the host galaxies as a function of observed redshift. The data in black and red points are collected from \cite{Sharma2024} and \cite{Gordon2023_1} respectively. In the right panel, we plot the distribution of stellar mass along with an approximate Gaussian function which roughly captures the observed distribution. We use this Gaussian approximate distribution in our calculation throughout this work.     }
\label{fig:stellarmass_dist}
\end{figure}

%----------------------------------------------
We use the sample of 53 FRBs provided in Table \ref{tab:FRB_sample} to obtain the observed distribution of stellar mass of their host galaxies. In the left panel of Fig. \ref{fig:stellarmass_dist}, we plot the stellar mass $M_*$ as a function of redshift. We see that the average stellar mass of host galaxies is about $10^{10}$ $M_{\odot}$. Most of the observed FRBs are at $z\lesssim 0.5$. Therefore, it is not possible to conclude how the distribution of stellar mass behaves at higher redshifts. In the right panel, we plot the probability distribution of $M_*$. We see that it is approximately normally distributed in ${\rm log}M_*$. We plot a Gaussian distribution with mean $\mu$=10 and standard deviation $\sigma=0.6$ to show the similarity between the two distribution. For simplicity, in this work, we use this approximate Gaussian distribution in order to capture the stellar mass function of potential FRB host galaxies.

%-----------------------------------------------

\end{document}